%% file: main.tex
\documentclass[journal]{IEEEtran}
 \usepackage{cite}
\usepackage[switch]{lineno}
 \usepackage{amsmath,amssymb,amsfonts}
 \usepackage{bm}
 \usepackage{graphicx}
 \usepackage{textcomp,nicefrac}
 \def\BibTeX{{\rm B\kern-.05em{\sc i\kern-.025em b}\kern-.08em
 T\kern-.1667em\lower.7ex\hbox{E}\kern-.125emX}}
\usepackage{siunitx}
\usepackage{color}
\usepackage{array,mathtools,cuted}
\usepackage{url}
\newcommand{\rd}[1]{\mathrm{d}#1}
\DeclareMathOperator{\erf}{erf}

\usepackage{xcolor}

\usepackage{tikz}
\usetikzlibrary{shapes.geometric, arrows}
\tikzstyle{data} = [ellipse, minimum width=3cm, minimum height=0.8cm,
    text centered, draw=black, fill=blue!20, font=\small, inner ysep=1.2pt]
\tikzstyle{process} = [rectangle, minimum width=4cm, minimum height=0.8cm,
    text centered, draw=black, fill=orange!20, font=\small, inner ysep=1.2pt]
\tikzstyle{arrow} = [thick,->,>=stealth]
\usetikzlibrary{calc}

\begin{document}

\title{GPU-Accelerated Analytic Simulation of Sparse Ionization Signal Formation in Pixelated Time Projection Detector}

% \author{
% Yousen~Zhang\IEEEauthorrefmark{1}\IEEEauthorrefmark{3},
% Brett~Viren\IEEEauthorrefmark{1},
% Mary~Bishai\IEEEauthorrefmark{1},
% Sergey~Martynenko\IEEEauthorrefmark{1},
% Xin~Qian\IEEEauthorrefmark{1}, \\
% Rado~Razakamiandra\IEEEauthorrefmark{1},
% Brooke~Russell\IEEEauthorrefmark{2} \\  \vspace{0.5em}
% \IEEEauthorblockA{\IEEEauthorrefmark{1}
% Brookhaven National Laboratory, Upton, NY 11973, USA
% } \\
% \IEEEauthorblockA{\IEEEauthorrefmark{2}
% Massachusetts Institute of Technology, Cambridge, MA 02139, USA
% } \\
% \IEEEauthorblockA{\IEEEauthorrefmark{3}
% Corresponding author: yzhang11@bnl.gov 
% }
% \comment{use your personal email instead?}
% }
\author{Yousen Zhang, Brett Viren, Mary Bishai, Sergey Martynenko, Xin Qian, Rado Razakamiandra, Brooke Russell
\thanks{
  This work was supported in part by the U.S. Department of Energy (DOE),
Office of Science, Office of High Energy Physics, under Contract No.~DE-SC0012704. (\textit{Corresponding author: Yousen Zhang}).

Yousen Zhang, Brett Viren, Mary Bishai, Sergey Martynenko, and Xin Qian are with the Physics Department,
Brookhaven National Laboratory, Upton, NY 11973 USA (e-mail: yzhang11@bnl.gov).

Xin Qian is also with the Department of Physics and Astronomy, Stony Brook University,
Stony Brook, NY 11794 USA.

Rado Razakamiandra is with the Department of Physics and Astronomy,
Stony Brook University, Stony Brook, NY 11794 USA.

Brooke Russell is with the Department of Physics,
Massachusetts Institute of Technology, Cambridge, MA 02139 USA.
}
  }

\maketitle

\begin{abstract}
  This paper presents a GPU-accelerated simulation package,
  \texttt{TRED}, for next-generation neutrino detectors with pixelated
  charge readout, leveraging community-driven software ecosystems to
  ensure adaptability and extensibility. 
  We introduce two generic
  contributions: (i) an effective-charge representation based on Gaussian
  quadrature rules, in which the linear-interpolation factors for the
  field response inside each voxel are absorbed into the effective
  charge, and (ii) a sparse,
  block-binned tensor representation that enables efficient FFT-based
  computation of induced signals on readout electrodes for sparsely
  activated detector volumes. The former captures structure inside a voxel
  without dense sampling, while the latter achieves low
  memory usage and scalable runtime, as demonstrated in benchmark
  studies. The underlying data representation is applicable to
  large-scale detectors and to other computational problems involving
  sparse activity.
\end{abstract}


%% internal review 
%% \linenumbers

\section{Introduction}
\IEEEPARstart{T}{he} Deep Underground Neutrino Experiment (DUNE) designs, constructs, and operates both the near-detector complex at Fermilab
and the far detector systems at the Sanford Underground Research Facility,
located approximately 1,300~km downstream~\cite{DUNE:2020lwj,DUNE:2021tad}.
One component of the DUNE near-detector complex is a liquid-argon time-projection chamber (ND-LAr), which employs the same target nucleus
and detector technology as the far detector to reduce detector-related systematic uncertainties~\cite{DUNE:2021tad}.
LArTPCs are fully active, homogeneous detectors that record millimeter-scale tomographic images of ionization charge together with prompt scintillation light,
enabling precise three-dimensional event reconstruction.
Owing to the high-power, wide-band neutrino beam provided by the Long-Baseline Neutrino Facility,
initially operating at 1.2 MW and upgradable to 2.4 MW, 
the near site experiences $\mathcal{O}(50)$ neutrino–argon
interactions within each $\sim10\,\mu$s beam spill,
delivered at a repetition rate of about 1 Hz.
Additional activity arises from muons generated in neutrino interactions
along the beamline upstream of ND-LAr and entering the detector volume,
with cosmic-ray-induced backgrounds negligible in comparison.
This high interaction rate results in dense, highly overlapping activity and imposes stringent requirements on detector readout,
motivating a modular LArTPC design with pixelated charge readout capable of resolving localized charge deposits along particle trajectories.

The ND-LAr detector comprises an active volume of approximately $7 \times 5 \times 3~\mathrm{m}^3$ distributed
across 35 TPC modules with millimeter-scale spatial sampling, resulting in $\mathcal{O}(5\times10^{5})$ readout channels per module.
Simulation and signal-processing software for LArTPC detectors has been under active development for more than a decade,
with established frameworks such as \texttt{Wire-Cell}~\cite{MicroBooNE:2018swd} 
providing mature detector-simulation and signal-processing capabilities.
These frameworks have enabled detailed detector modeling and reconstruction workflows based on channel-by-channel simulation.
Scaling to the channel counts and sparsity patterns characteristic of ND-LAr remains challenging
in CPU-based implementations, which can require substantial core counts and computing resources.
More recently, GPU-accelerated approaches have demonstrated the feasibility of
highly parallelized simulations for pixelated LArTPCs~\cite{DUNE:2022gxa},
including designs that scale with sparse, localized charge activity along track trajectories.
Nevertheless, such approaches can still place substantial demands on GPU resources
because of the underlying data representation and computational architecture.
In particular, memory capacity and bandwidth may become limiting factors at large detector scale.
These considerations motivated the design developed in this work.
It aims to improve resource efficiency and scalability by
using a block-sparse representation tailored to localized activity,
enabling efficient processing of large detector volumes while retaining high fidelity.

In this work, we present a GPU-native simulation package, \texttt{TRED}~\cite{tred},
built on community-driven software ecosystems,
exemplified by \texttt{PyTorch}~\cite{paszke2019pytorch},
to ensure long-term adaptability and extensibility,
and tailored to the highly sparse and localized data characteristic of LArTPC detectors.
The \texttt{PyTorch} backend provides access to ongoing support for GPUs
and modern computing architectures developed by
the broader computer science and machine learning communities.
In addition, \texttt{TRED} is distributed as a standard Python package via 
\texttt{pyproject.toml}~\cite{pep621} and can be installed using common tools such as \texttt{pip}, \texttt{uv}, and \texttt{poetry}.

We introduce a high-fidelity effective-charge representation of the ionization distribution
after recombination and transport, together with a block-sparse binned tensor data structure
that preserves locality while enabling efficient batched execution on GPUs.
The induced signal on electrodes admits an analytic expression as a convolution of the effective-charge
distribution with the detector Green’s function.
The block-sparse binned representation enables this convolution to be evaluated
using FFT-based methods on sparse charge distributions, without constructing a
global dense grid.

Although developed with the DUNE ND-LAr detector in mind,
the underlying concepts are applicable to a wider class of large-volume detectors
with sparse activity, including the simulation of ionization charge signals
in projective-readout LArTPCs such as the DUNE far detector.
The block-based sparse data representation further supports broader applicability,
including sparse sub-manifold convolution networks~\cite{graham2017submanifold} used in machine learning.

Section~\ref{sec:overview_simulation} discusses the simulated detector physics in LArTPCs.
Section~\ref{sec:software} describes the software design and implementation of \texttt{TRED}.
Section~\ref{sec:performance} presents performance measurements,
resource scaling, and associated limitations.
Section~\ref{sec:discussion} discusses these results in a broader physics and computing context.
We summarize our results and outline future directions in Sec.~\ref{sec:summary}.

\input{simulation}

\section{Software design}~\label{sec:software}
\subsection{Array-Oriented, Graph-Based Architecture}

Modern GPU architectures are optimized for data-parallel workloads operating on contiguous memory layouts.
Accordingly, we adopt an array-oriented design in which the primary data structure is an N-dimensional tensor.
Core operations, such as Hadamard (element-wise) products and convolutions,
are expressed in idiomatic array form and executed using highly optimized backend libraries.
This design aligns naturally with established scientific computing frameworks,
and \texttt{PyTorch} is used as the backend due to its mature GPU support and extensive user community.

To structure the computation, we adopt a graph–based architecture
in which each operation is encapsulated as a \texttt{PyTorch} \texttt{nn.Module} (a graph-node)
with well-defined inputs and outputs.
These modules form a directed acyclic graph that mirrors the simulation pipeline, shown in Fig.~\ref{fig:flowchart}.
By packaging computational kernels as modules, we delegate kernel dispatch,
device and stream placement, and memory management to the PyTorch runtime,
while naturally enabling practical capabilities such as automatic batching, mixed precision,
and profiler-guided optimization.
The module abstraction provides a uniform interface that substantially improves composability,
testability, and long-term maintainability.

\subsubsection{Effective charge calculation}
The expression for $Q_{ijk}^{\text{eff}}$ from Eq.~\eqref{eq:eff_q} can be interpreted as an element-wise product with quadrature weights, followed by convolution with trilinear coefficients.
The quadrature weights are mapped from the standard interval $[-1,1]$ to the local intervals surrounding each grid point.
The trilinear coefficients $u$ depend only on the quadrature node positions within the standard interval, and therefore, can be precomputed.
The resulting formulation closely resembles a three-dimensional convolution
(\texttt{Conv3D}) with a small-support kernel $u$, enabling particularly
efficient evaluation on GPU architectures.

\subsubsection{Mirror-Pair Complex Packing (MPCP) Convolution}
\label{sec:mpcp}
As noted in Sec.~\ref{sec:det}, the induced current on the electrodes arises from the convolution between the effective charge distribution and the field response function. This is realized by partitioning the entire space into sub-grids according to the impact positions of electrons relative to the electrode centers. In machine learning terminology, this corresponds to a channel-wise convolution followed by a summation over channels. Partitioning the input prior to convolution is equivalent to array slicing in array-oriented languages. To accelerate the computation with a large kernel (the field response), we employed an FFT-based linear convolution technique.
It is also important to note that the field response exhibits reflection symmetry with respect to the electrode’s center.
We employ a method, named ``Mirror-Pair Complex Packing (MPCP)'', to further boost the calculation.

To illustrate this, consider an one-dimensional case. For each sub-electrode offset \( j \), the response observed at the \( i \)-th electrode is denoted as \( R_i^j \). The induced current at electrode position \( p \) is then given by
\[
I_p = \sum_{i,j} Q_i^j R_{p-i}^j ,
\]
where each charge element \( Q_i^j \) contributes according to the offset \( p-i \) from its position. Under reflection symmetry, the response satisfies
\[
R_{-i}^{-j} \equiv R_i^j .
\]
Using this condition, the induced current can be written as
\begin{equation}
  \label{eq:multch}
  \begin{aligned}
I_{p} &= \sum_{j=-N,\, j \neq 0 }^{N} \sum_i R_{i}^{j} Q_{p-i}^j \\
      &= \sum_{j=1}^{N}\sum_{i} R_{i}^{j}\bigl(Q_{p-i}^j + Q_{p+i}^{-j}\bigr) \\
    &= \sum_{j=1}^{N}\sum_{i} R_{i}^{j}\left(Q_{p-i}^{j} + Q_{-(p'-i)}^j \right), \\
    \end{aligned}
\end{equation}
where $p'=-p$. %The last term encodes two reflections: the leading minus in $-(p'-i)$ flips the input $Q$ along the index $i$, and the relation $p'=-p$ flips the output (the induced current indexed by $p$).
The second term corresponds to a reflection in both the input index and the output position.

Although the ionization terms do not combine into a single convolution because of the absence of reflection symmetry in the ionization charge, the two contributions can still be evaluated efficiently.  Note that the pair of channels can be packed into a single complex array, $Q^{j} + \mathrm{i}\,Q^{-j}$ (with $\mathrm{i}^2=-1$), so they are transformed simultaneously and do not interfere during the FFT operations. This complex packing reduces the number of FFTs. 

It is worth noting that the multi-channel convolution can be evaluated efficiently in the frequency domain:
\begin{equation}
  \begin{aligned}
I_{p}
&= \sum_{j}\operatorname{IFFT}\!\left(\operatorname{FFT}(R^{j})\,\operatorname{FFT}(Q^{j})\right) \\
&= \operatorname{IFFT}\!\left(\sum_{j}\operatorname{FFT}(R^{j})\,\operatorname{FFT}(Q^{j})\right).
\end{aligned}
\end{equation}
Consequently, with precomputed $\operatorname{FFT}(R^j)$, each
evaluation requires one forward FFT per (packed) impact position and a
single global inverse FFT; MPCP further halves the number of forward
FFTs.

\subsection{Resources management}
\subsubsection{Block sparse binned tensor}
\label{sec:bsb}
A LArTPC with pixelated readout comprises thousands of readout channels and a large number of time samples along the drift direction.
At the same time, neutrino interactions produce ionization signals
that occupy only a small fraction of the full active detector volume
and activate a limited subset of readout channels at any given time,
even for the longest muon tracks from charged-current interactions.
The combination of a large detector readout space and intrinsically sparse signal activity makes it impractical and computationally inefficient
to track the induced current on every pixel at every time sample.
An efficient representation therefore requires the use of sparse data structures.

We exploit sparsity by implementing a concept called block-sparse binned tensor.
In this scheme, dense data are stored within blocks, where each block is composed of an N-dimensional array and an accompanying length-N vector.
The array stores the physical values at grid points,
while the vector specifies the coordinates of the block’s lower corner.
Blocks have a regular shape and can be batched for GPU processing.
The entire detector volume (or a sub-volume of interest) is partitioned into such bins,
but only those containing event activity are retained.
Importantly, bin positions are required to align with the bin boundaries of the enclosing volume.

Beyond storage efficiency, the block-sparse binned tensor provides additional functionalities for simulation.
First, it allows input arrays to be aligned and cropped to bin boundaries.
Second, it handles cases where diffusion and convolution of field response function enlarge the data footprint,
causing originally disjoint inputs to produce overlapping outputs. 
This overlap can create contention during subsequent parallel operations.
The block-sparse binned tensor resolves this issue by reducing overlapping regions
(using summation or other reduction operations) implemented efficiently through PyTorch’s \texttt{scatter\_add} backend.
The concept of the block-sparse binned tensor is analogous to the histogram described
in Ref.~\cite{Ahn:2005sa}.
However, in this work, all data are organized and processed in binned blocks,
requiring additional mapping and reduction operations between high-dimensional tensors,
rather than simple scalar additions.

\subsubsection{Batching and chunking for sparse data}
\label{sec:chunk}

Fixed batch sizes are commonly used in traditional workflows, but they are poorly suited to sparse signals:
localized bursts of activity can produce transient peaks in memory usage that undermine predictable resource control.
To address this, we replace globally fixed batching and batch-only chunking with two complementary strategies.
Hierarchical batching dynamically adjusts the batch size to the realized sparsity and computational load,
while sub-dimensional chunking partitions tensors along orthogonal spatial or temporal axes,
such as the electron drift direction, to bound memory consumption without sacrificing throughput.
As an example, during rasterization of \texttt{GEANT4} track segments onto a spatial grid (described in Sec.~\ref{sec:det}),
different segments intersect different numbers of grid cells, resulting in a variable number of rasterized blocks even for fixed-size inputs.
To accommodate this variability while maintaining efficient downstream execution, we perform a re-batching step that reorganizes the rasterized blocks into batches of appropriate size.
This re-batching groups blocks according to their realized occupancy and workload, ensuring balanced computation and bounded memory usage in subsequent processing stages.

Hierarchical batching also applies in cases where downstream operations increase tensor dimensionality and require additive reductions.
In LArTPC simulations, both rasterization and field-response convolution expand the spatial support of the signal,
leading to overlapping contributions that must be accumulated before subsequent processing steps.
Because these reductions are linear, the ordering of batching operations does not affect the final result to a noticeable level.

Furthermore, evaluating the effective charge using quadrature introduces additional node axes along each dimension,
increasing the data volume multiplicatively with the number of quadrature nodes.
As a result, even modest batch sizes can become memory intensive.
To control memory usage, we partition the computation by chunking tensors along orthogonal axes,
% specifically along feature dimensions that are independent in the convolution,
such as the electron drift direction.
Each chunk is processed independently and sequentially, bounding peak memory consumption while preserving overall throughput.
% The required chunking and subsequent re-batching introduce negligible overhead compared to the cost of block-wise GPU computation.
% Because \texttt{GEANT4} employs dynamic stepping, we propose hierarchy batching and dynamic sub-dimensional chunking that adapt to input size,
% better accommodating sparsity and reducing memory consumption.

A summary of control flow is illustrated in Fig.~\ref{fig:flowchart}. The augmented features from the global batched input are divided into smaller batches $i$ during the rasterization process. Each rasterized batch is then further split into sub-batches $j$, which are processed through convolution. The resulting outputs are binned and accumulated, as described in Sec.~\ref{sec:bsb}, to obtain the induced current at unique pixel locations for each $(i, j)$ pair. For each rasterized batch $i$, the currents are summed over the unique pixels. After iterating over all $i$-batches, a final concatenation and accumulation step is performed to combine the contributions from the entire global input.

\section{Performance evaluation}~\label{sec:performance}
\subsection{Test data and simulation setup}
\begin{figure}
\centering
\includegraphics[width=0.9\linewidth]{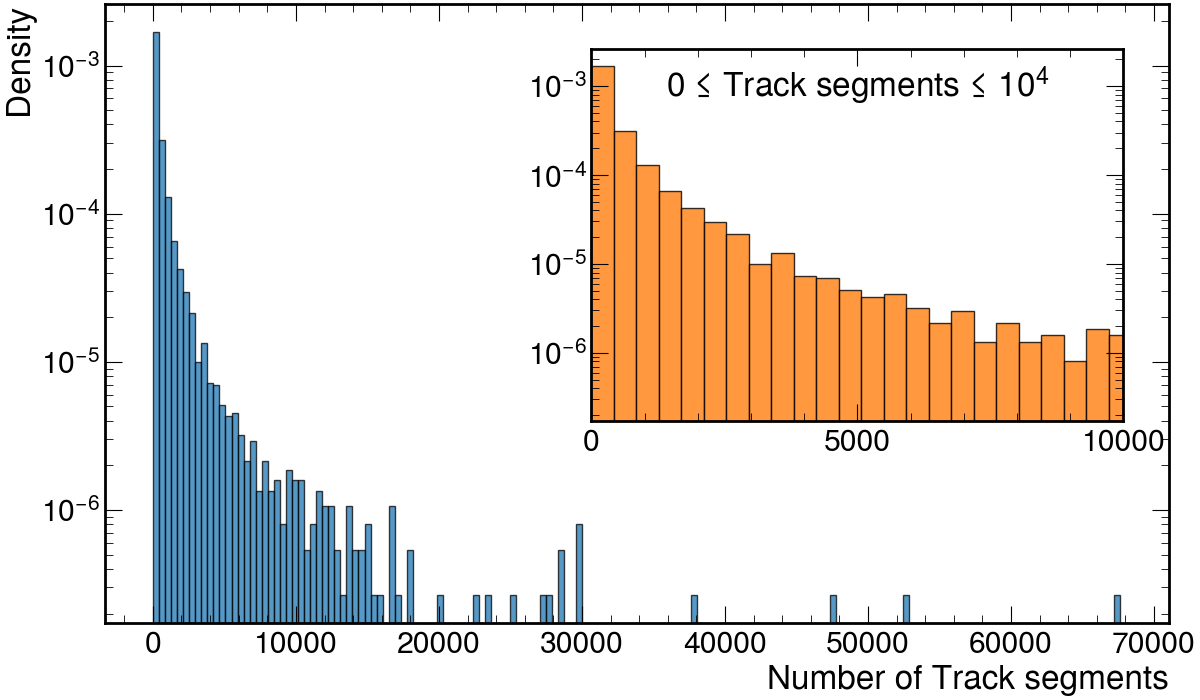}
\caption{Distribution of the number of track segments per neutrino beam spill and per TPC drift volume,
simulated from \texttt{edep-sim}. 
    \label{fig:DistNsegs}}
\end{figure}

The data sample used in this study corresponds to the anti-neutrino beam configuration at the DUNE-ND site.
The neutrino-argon interaction cross section in the LArTPC
are simulated using the \texttt{GENIE} event generator~\cite{Andreopoulos:2015wxa}.
On average, about $\mathcal{O}(50)$ neutrino interaction vertices occur
within ND-LAr for each spill of the primary proton beam across 35 dual TPCs.
The \texttt{GENIE} outputs are then passed to \texttt{edep-sim}~\cite{McGraw:2021a,GEANT4:2002zbu},
a \texttt{GEANT4}-based simulation framework, to estimate the energy depositions
from the products of neutrino-argon interactions inside the detector.
The ND-LAr detector consists of $7 \times 5$ TPC modules, each comprising
two drift volumes separated by a central cathode.
Each drift volume has a maximum drift length of approximately 50 cm and is instrumented with a pixelated anode plane.
Each anode plane is approximately 1 m long in the neutrino beam direction and 3 m tall.
We partition the step segments according to their containing TPC drift volumes and process these volumes independently in sequence.
Figure \ref{fig:DistNsegs} shows the distribution of input \texttt{G4Step} objects (track segments) per neutrino beam spill and per TPC drift volume, simulated with \texttt{GEANT4}.
We initially focus on cases with fewer than $\mathcal{O}(5000)$ $G4Step$ objects;
as demonstrated later, the performance of our approach scales readily to larger occupancies.
For the full-spill configuration considered here, the simulation can be processed with a memory footprint below 20~GB and a runtime below 10~minutes on an NVIDIA GeForce RTX 4090.
For reference, the same full spill requires more than 150~minutes on a single CPU thread of an AMD Ryzen Threadripper 7970X.

The field response is calculated on a $5\times 5$ pixel grid, and spanning approximately 15 cm along the nominal drift direction,
corresponding to about 2000 time ticks at 50 ns per tick.
We use a baseline setup where the area around an electrode is divided into $10\times 10$ subregions,
to sample the field response function. This resolution captures the relevant spatial features of the response.
This choice also aligns with the typical diffusion width,
which is comparable to or larger than the subgrid spacing in this configuration most of the time.
The geometry is illustrated in Fig.~\ref{fig:frillu}.
The readout model assumes the LArPix detector design, as described in Sec.~\ref{sec:readout}.
The preamplifier noise is modeled as an independent normal distribution with a standard deviation of 500 electrons.
The discriminator threshold is also affected by random noise, with a standard deviation of 650 electrons.
In addition, each reset introduces a random baseline shift.
The width of this baseline distribution is about 900 electrons.
These values are based on benchtop measurements provided by the ASIC team.

The rasterized charges are grouped into $4\times4$ pixels and $32$ time ticks before convolution.
The convolution output is then padded to $8\times8\times2048$ to maintain high performance by using power-of-two dimensions.
The choice of $32$ time ticks is motivated by the runtime study shown in Fig.~\ref{fig:runtimevsinput}.
The waveform input to the readout electronics model is configured to span $600 \mu\text{s}$ (approximately 96 cm),
corresponding to twice the maximum drift length of the DUNE-ND modules.
The tests were performed on an NVIDIA GeForce RTX 4090 GPU with 24 GB of memory using 64-bit float numbers.

\begin{figure}[htb]
\centering
\includegraphics[width=0.95\linewidth]{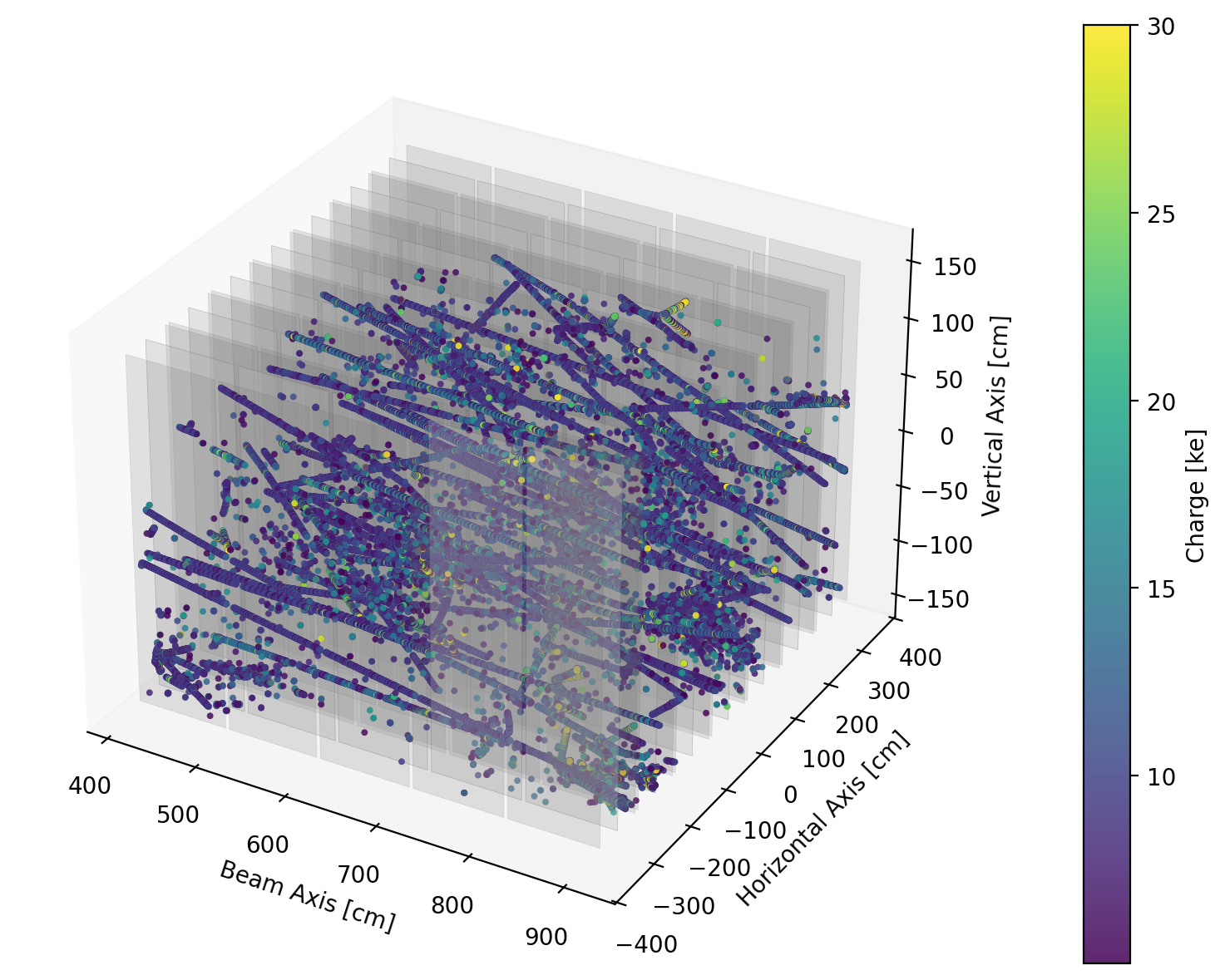}
\caption{Event display of reconstructed detector activity from a full beam-spill simulation in ND-LAr, including overlapping neutrino interactions and associated secondary activity under high pile-up conditions.}
\label{fig:event}
\end{figure}

The reconstructed detector activity from a full beam-spill simulation is shown in Fig.~\ref{fig:event}. The event display includes overlapping neutrino interactions occurring within ND-LAr as well as secondary particles originating from neutrino interactions upstream of the detector volume. The resulting topology illustrates the high-occupancy environment expected during a beam spill and demonstrates the capability of the simulation framework to model the full detector response under realistic pile-up conditions.

\subsection{Performance Benchmarks: Accuracy, GPU Memory Usage, and Runtime}
\begin{figure}[!htbp]
  \centering
  \includegraphics[width=0.98\linewidth]{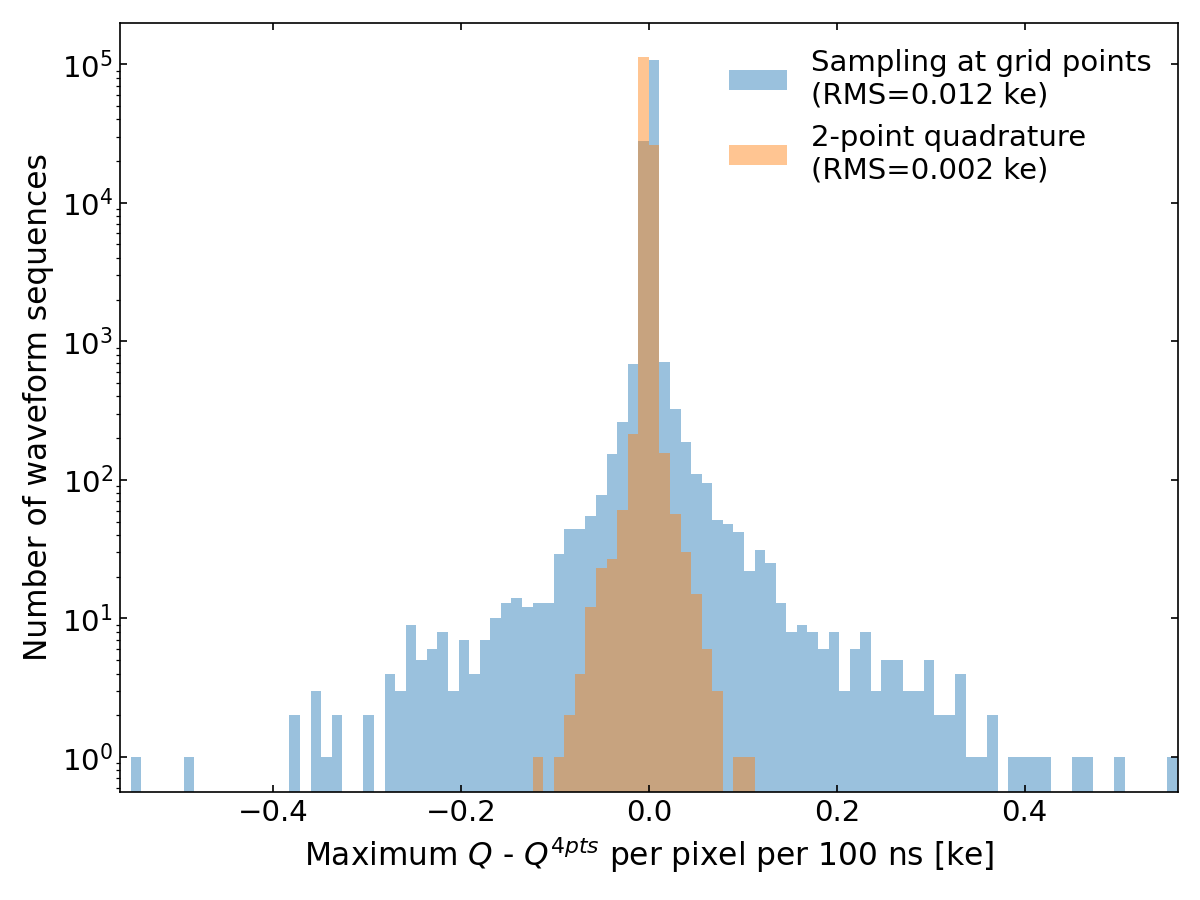}
  \caption{The maximum deviations of accumulated charge per pixel channel per 100ns, under sampling at grid points, or two-point quadrature rule, relative to the reference charge under 4-point quadrature rule.}
  \label{fig:accuracy}
\end{figure}

Figure~\ref{fig:accuracy} presents the accuracy study.
The maximum deviation of the accumulated charge per pixel channel within a 100\,ns window
is evaluated for normal sampling at grid points and for the two-point quadrature rule,
each relative to the reference obtained using the four-point quadrature rule.
Taking the higher-order quadrature as a reference,
the substantially narrower spread observed for the two-point quadrature
demonstrates that the effective-charge approach captures
the fine structure of the induced waveform without requiring a finely gridded field response function.
The corresponding RMS deviations are $12\,e^{-}$ for sampling at grid points
and $2\,e^{-}$ for the two-point quadrature rule, 
representing an improvement of approximately a factor of six.
These residual differences are substantially smaller than
the typical equivalent noise level of front-end electronics in LArTPCs, $\mathcal{O}(500\,e^{-})$.

\begin{figure}[!htbp]
    \centering
    \includegraphics[width=0.98\linewidth]{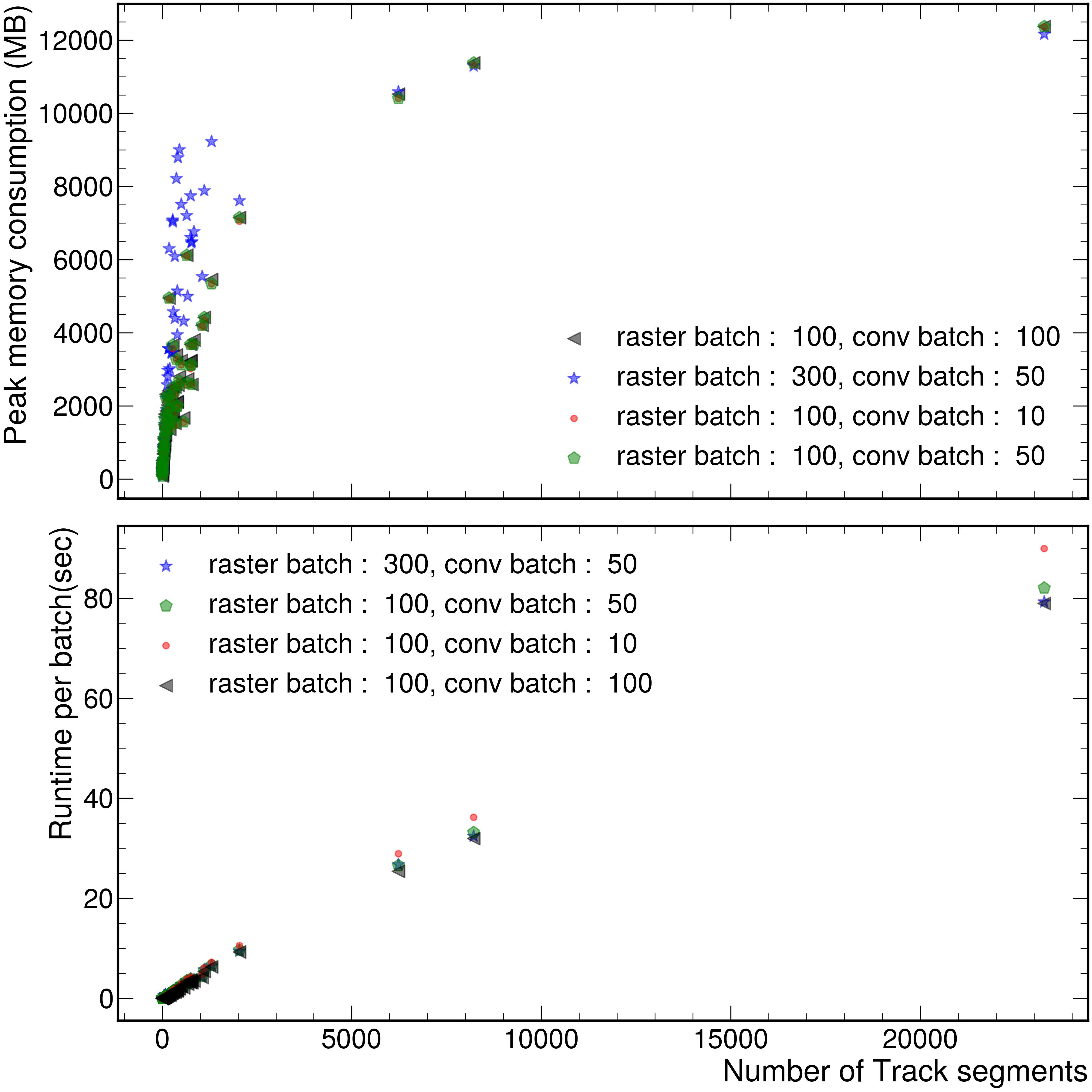}
    \caption{GPU memory usage (top) and runtime (bottom) per TPC drift volume as a function of the number of input G4Step objects {tracks
    segments} for different batching schemes. \textbf{Top:} GPU memory usage per batch when the hierarchy batching 
    is enabled, and the hierarchy chunking is disabled. \textbf{Bottom:} Runtime per batch when the hierarchy batching 
    and chunking are enabled.}
    \label{fig:runtimeGPUMem}
\end{figure}

\begin{figure}
   \centering
   \includegraphics[width=0.98\linewidth]{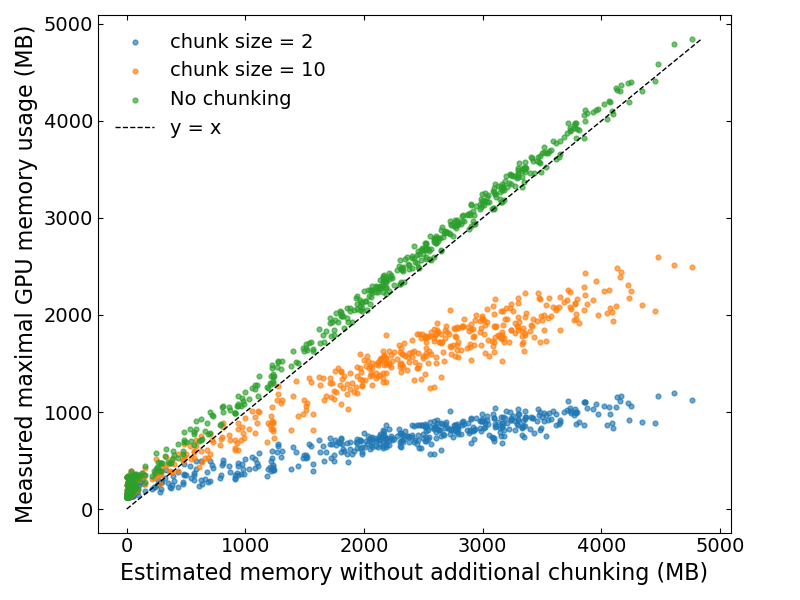}
   \caption{Memory usage with chunking over one axis in the pixel plane, shown as a function of the estimated memory usage without chunking over the feature dimensions.
     The dashed line indicates the ideal case where the measured memory usage coincides with the estimated value.
     Green points indicate the memory usage without chunking over the feature dimensions. Blue and orange points show the results for chunk sizes of 2 and 10, respectively. }
    \label{fig:chunkx}
\end{figure}

One key challenge in the GPU programming lies in managing memory usage.
The hierarchical batching and sub-dimensional chunking strategies described in Sec.~\ref{sec:chunk} address the
data volume increase introduced by applying both the quadrature rule and convolution.
The upper panel of Fig.~\ref{fig:runtimeGPUMem} shows the peak GPU memory usage with hierarchical batching enabled and chunking disabled.
For a rasterization batch size of $i=100$, the memory usage increases only gradually with activity, reflecting the slow growth of the output-dominated contribution.
As event activity increases, all configurations converge to the same asymptotic memory usage,
since the dominant memory footprint is determined by long per-pixel waveforms
that are independent of batching or chunking parameters.
In contrast, when the rasterization batch size is increased to $i=300$, a rapid rise in memory usage is observed at low activity,
because memory consumption is no longer dominated by the output tensors but instead by intermediate states generated during the rasterization step,
including the sparse representations of accumulated charge blocks.
These intermediate structures scale strongly with the rasterization batch size and therefore drive the observed memory increase.
GPU memory usage can be further reduced by chunking along feature dimensions, as described in Sec.\ref{sec:chunk}.
We evaluate this effect by applying different chunk sizes along a single axis in the pixel plane.
As shown in Fig.~\ref{fig:chunkx}, introducing chunking significantly lowers peak memory consumption compared with the unchunked case,
with reductions of approximately a factor of 2–5 depending on the chunk size.

We now examine its impact on runtime performance.
The lower panel of Fig.~\ref{fig:runtimeGPUMem} shows the corresponding GPU runtime per batch for different combinations
of raster batching (ionization-charge batching) and convolution batching.
For a fixed raster batch size, comparing different convolution batch sizes demonstrates
that the overhead introduced by splitting the computation into smaller batches is minimal.
A noticeable runtime increase of about 10\% is observed only when the convolution batch size is reduced to 10.
Beyond this point, the runtime scales approximately linearly with input size,
indicating that the GPU resources are fully utilized and that performance
is dominated by the computational workload rather than batching overhead.
The approximately linear relationship demonstrates that the software scales
well with the number of input step segments.

\begin{figure}
    \centering
    \includegraphics[width=0.9\linewidth]{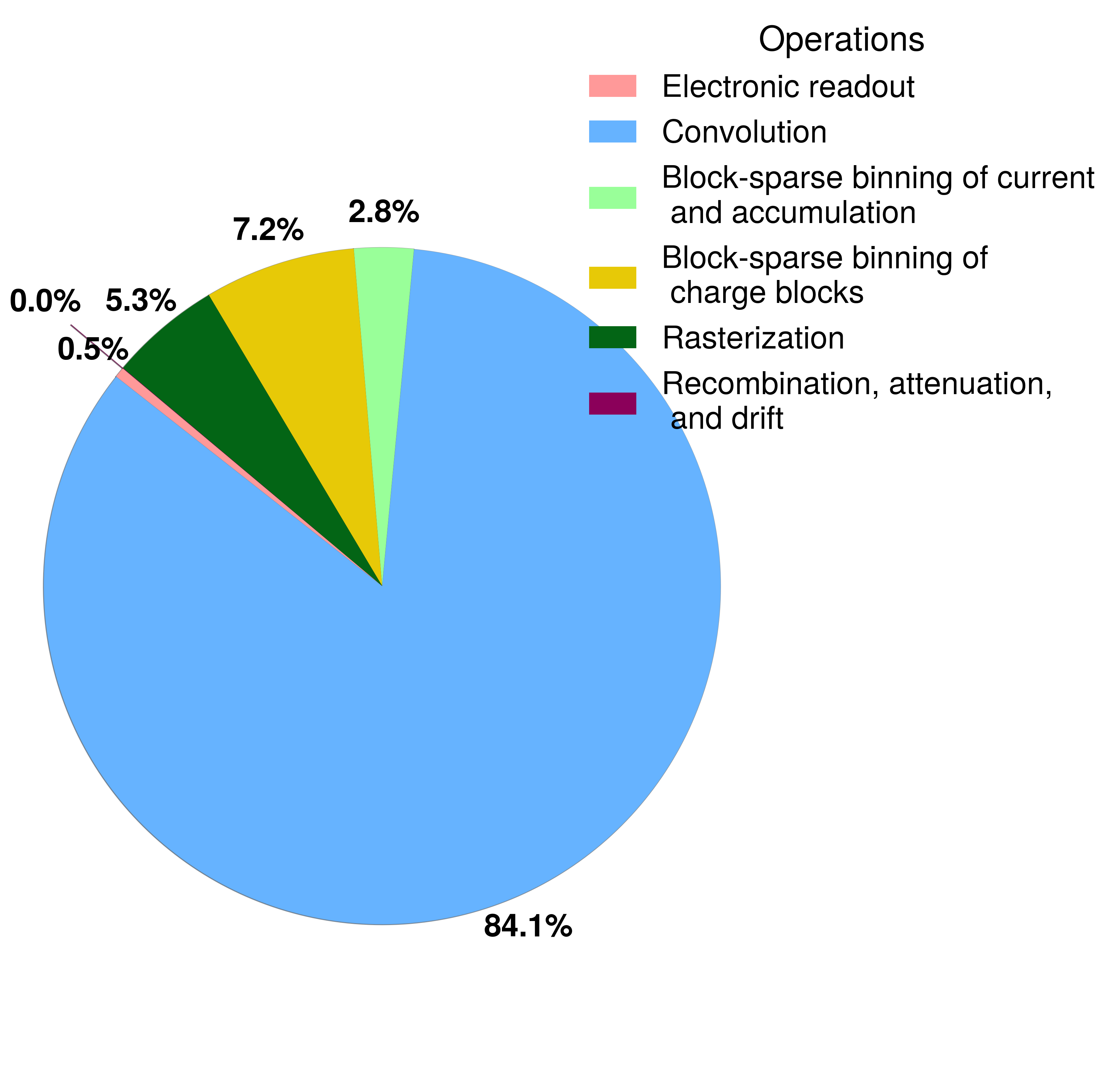}
    \caption{Runtime share among the major operations on the GPU. Each slice represents an operation's total runtime relative to the total execution time.}
    \label{fig:runtimepiechart}
\end{figure}

We analyze the runtime performance broken down by operation.
One batching scheme is selected since the differences among schemes are minor.
Figure~\ref{fig:runtimepiechart} shows the runtime distribution among the major operations aggregated across all TPCs.
The large slice for the convolution indicates that the convolution dominates the overall execution time and constitutes the principal performance bottleneck.
The block-sparse binning of the charge blocks is the second most time-consuming operation.

\begin{figure}
    \centering
    \includegraphics[width=0.9\linewidth]{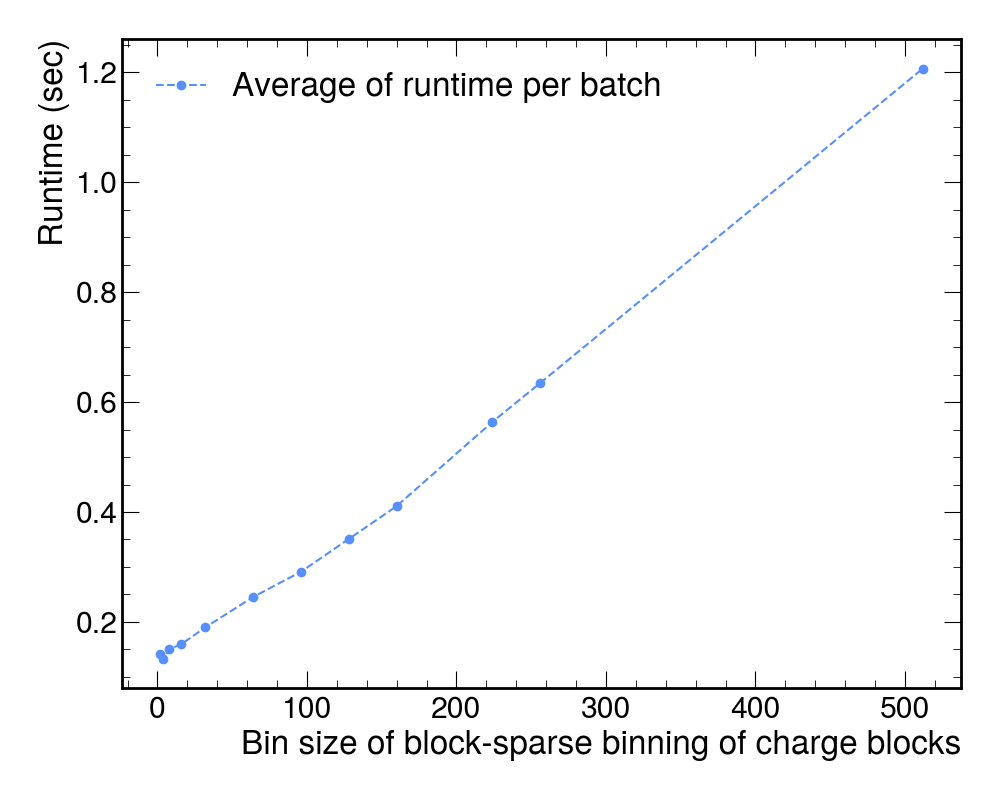}
    \caption{
      Runtime for converting blocks into a block-sparse binning representation as a function of bin size.
      Each point shows the average time for the block-sparse binning operation applied to the same charge blocks.}
    \label{fig:runtimevsinput}
\end{figure}

We further evaluate the runtime performance for different choices of bin size in the block-sparse binned tensor representation.
Figure~\ref{fig:runtimevsinput} shows the GPU runtime of the block-sparse binning operation
applied to the charge blocks as a function of the bin size along the time axis. As the bin size increases, the runtime also increases.
This behavior arises because larger bins tend to include a higher fraction of zero entries, reducing the effectiveness
of sparsity and increasing unnecessary computation.
In the temporal direction, the typical extent of ionization-charge segments ranges from approximately 50 to 150 ticks.
For bin sizes below about 50 ticks, the runtime remains both fast and stable,
indicating that sparsity is effectively preserved in this regime. As the bin size increases beyond this scale,
the performance degrades as bins increasingly encompass regions without signal.

These observations indicate that optimal performance is achieved by choosing a bin size smaller than the characteristic
size of the input charge blocks. However, this choice involves a trade-off, since the block-sparse binned tensor serves as the input to subsequent operations.
For example, when used as input to a convolution, bins that are much smaller than the field response increase the number of convolution operations, while the output size changes only marginally. As a result, excessively small bin sizes can lead to increased overall runtime despite faster binning. An optimal bin size therefore balances efficient sparse binning against the computational cost of downstream operations.

\section{Discussions}~\label{sec:discussion}
We have presented the performance of the \texttt{TRED} package and demonstrated
that its analytical treatment of effective charge enables accurate and efficient induced-current calculations.
% By combining weighted quadrature rules with trilinear interpolation of the field response,
% \texttt{TRED} improves numerical accuracy, while the use of FFT-based linear convolution ensures scalable performance on modern computing architectures.
In this section, we discuss key aspects of the effective-charge formulation and software architecture,
examine its suitability for calibration and differentiable simulation workflows,
and outline potential extensions of the approach to other detector configurations, including the DUNE far detector.

\subsection{Accuracy, Applicability, and Numerical Considerations of Charge Calculation}
The accuracy of the quadrature-based formulation depends on the relative scale of the integration intervals
and the diffusion width of the charge distribution.
When the grid spacing is significantly larger than the diffusion width, the approximation degrades,
as the charge distribution is insufficiently resolved. In practice, this situation may arise either
when diffusion is negligibly small or when the simulation grid is chosen too coarsely.
The former is uncommon in large liquid-argon TPCs and similar neutrino detectors,
where long drift distances naturally lead to substantial diffusion,
and can be further mitigated by imposing a minimum diffusion scale in the simulation.
The latter can be readily avoided by selecting a grid spacing commensurate with the expected diffusion scale, a choice that remains under user control.

More generally, many simulation frameworks for projective readout detectors treat individual energy depositions
as point-like objects prior to diffusion. The effective-charge formalism introduced here remains applicable in such cases,
provided a suitable criterion is adopted to transition between point-like and extended (line-like) representations.
This flexibility allows the method to be used consistently across different detector models and simulation paradigms.

We also note that mapping interpolation coefficients from the field response to the charge density at quadrature nodes
can introduce mild smearing near pixel boundaries.
This arises because the transition from unipolar signals in the central collection pixel
to bipolar currents in neighboring pixels occurs over a very short spatial scale.
Interpolation across this sharp transition can smooth the response and lead to a small spreading of charge across pixel boundaries.
In practice, this effect is confined to a narrow region near pixel edges and becomes negligible when the grid spacing is sufficiently fine,
as the fraction of affected grid points scales with the boundary area rather than the pixel volume.
Moreover, the smearing can be corrected if needed by applying boundary-aware interpolation or post-processing corrections.
In our tests, the effect is already strongly suppressed due to the use of fine spatial grids and continuous charge distributions,
and any residual impact is negligible, shown in Fig.~\ref{fig:accuracy}.

\subsection{Software Abstractions and Broader Applicability}

The package presented in this work emphasizes abstraction and deliberately leverages concepts and tools from the broader computational science community—particularly those developed for machine learning to enhance long-term maintainability and extensibility. This design choice also provides a potential pathway toward portability across diverse hardware architectures supported by \texttt{PyTorch} backends, as these ecosystems continue to mature. The extent to which such portability can be realized while maintaining competitive performance remains an important subject for future investigation.

The block-sparse binned tensor introduced here further serves as an effective abstraction for computational problems that require careful coordination of concurrent memory reads and writes, such as those encountered in sparse neural network implementations. Many existing frameworks either provide limited support for these access patterns or are restricted to two-dimensional sparse tensors. For example, \texttt{SparseConvNet} and \texttt{MinkowskiEngine} represent multi-dimensional sparsity using coordinate (COO) formats, which are not well suited for FFT-based operations. In contrast, the block-sparse binned tensor formulation naturally supports structured sparsity in higher dimensions while remaining compatible with convolutional and FFT-based workflows.

\subsection{Differentiable Simulation and Calibration Potential}

The \texttt{PyTorch}-based implementation also enables end-to-end simulation and optimization by directly comparing simulated outputs with data from calibration sources~\cite{Gasiorowski:2024a}. Unlike traditional calibration workflows in particle physics, which typically adjust parameters sequentially and in isolation, an end-to-end approach allows all relevant parameters to be optimized simultaneously within a unified framework. This can improve internal consistency between simulation components and enhance the efficiency and accuracy of detector modeling.

Within our framework, the input parameters of individual computational layers can,
in principle, be optimized automatically using output data, forming a differentiable
and self-consistent calibration pipeline.

As a proof of concept, we implemented a variant of the main package
and used it to tune the electron lifetime under the assumption that the simulated waveforms,
their corresponding reference waveforms, and source information for the test samples are available.
The reference waveform includes a realistic readout-noise level of 500 ENC per time tick.

\begin{figure}[ht]
  \centering
  \includegraphics[width=0.95\linewidth]{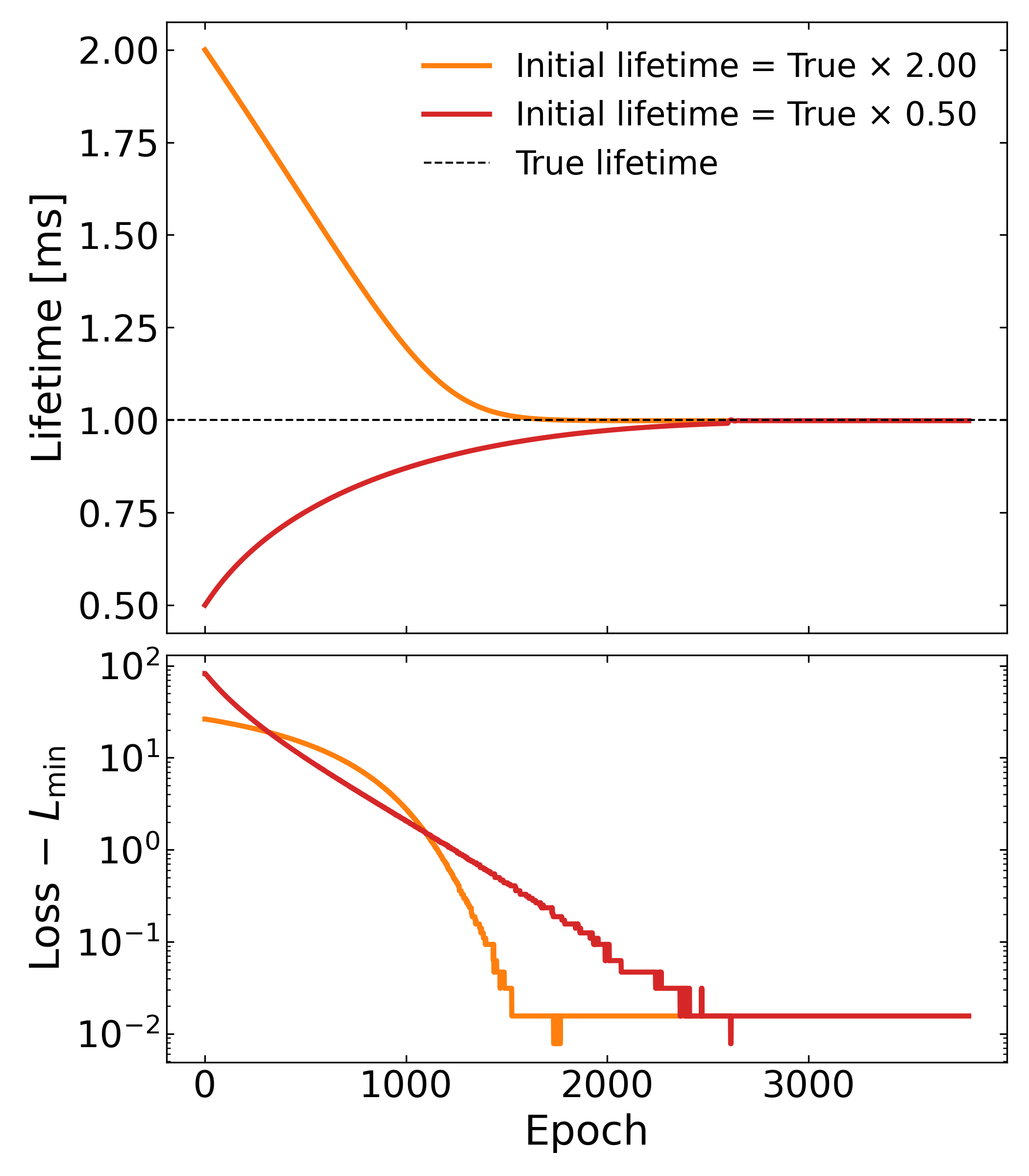}
\caption{Upper panel: fitted electron lifetime as a function of training epoch (iteration). Lower panel: loss relative to the minimum loss value. The dashed line indicates the true lifetime, and the solid lines show the optimization trajectories starting from initial values of 2.0$\times$ and 0.5$\times$ the true lifetime. The reference waveform includes a readout-noise level of 500 ENC per time tick.}
  \label{fig:differentiable}
  \end{figure}

The results are shown in Fig.~\ref{fig:differentiable}. The true electron lifetime is set to 1.0 ms,
and two initial values are chosen to be 2.0 and 0.5 times the true value.
The upper panel shows the evolution of the fitted lifetime during optimization,
while the lower panel shows the loss relative to its minimum value.
Because the reference waveform contains noise, the loss values are evaluated with respect to
a common noise-induced baseline rather than converging to zero.
Even with this noise and the deliberately displaced initial guesses,
the optimization converges toward the true lifetime in both cases,
illustrating stable gradient-based tuning in this proof-of-concept setting.

We emphasize, however, that automatic calibration is not a core design goal of the framework. Gradient-based optimization requires intermediate states to be retained until the end of each optimization epoch, which limits memory reuse and can significantly increase memory consumption. 
As a result, while the framework supports differentiable simulation and optimization, its primary focus remains fast and accurate forward simulation rather than automated calibration.

\subsection{Relation to Existing ND Simulation Tools}
For the DUNE Near Detector, the existing simulation package \texttt{larnd-sim}\cite{DUNE:2022gxa}
employs custom GPU kernels with just-in-time compilation via \texttt{Numba}\cite{Lam:2015a},
managing data in raw arrays and enforcing sparsity through pixel-level operations.
While both frameworks simulate the same underlying detector physics,
they differ substantially in software architecture, abstraction level,
and use of community-driven ecosystems.
We find that \texttt{TRED} reproduces
the induced-current waveform predicted by \texttt{larnd-sim}.
Comparisons with detector data also show that
key signal features observed in the data are reproduced by the simulation.
A detailed, quantitative comparison of performance, resource usage,
and physics validation between \texttt{TRED}
and existing ND simulation frameworks,
including comprehensive comparisons with experimental data,
is deferred to future work.

\subsection{Extension to Far-Detector Simulations}

The \texttt{TRED} package was developed with a primary focus on the DUNE Near Detector, which employs zero-suppressed front-end electronics. Nevertheless, the underlying functions and tools are readily extendable to continuously sampled readout detectors, such as the DUNE Far Detector, by replacing the field-response–only model with a combined field-response and electronics-response formulation. Because the response functions are precomputed, this extension introduces no additional computational overhead during simulation.

As a proof of concept, test cases in the cosmic-ray environment
of the ProtoDUNE Single-Phase detector~\cite{DUNE:2017pqt},
simulated using \texttt{Wire-Cell}~\cite{MicroBooNE:2018swd,wct},
demonstrate a speed improvement of approximately a factor of two on CPU architectures.
This performance gain highlights the effectiveness of the sparse data representation employed by \texttt{TRED}.
In contrast, the standard \texttt{Wire-Cell} workflow defines a global region of interest for each drift volume
and performs convolution on a dense representation of the signal using a detector response function that combines both field and electronics effects. 

The advantages of \texttt{TRED} become even more pronounced for far-detector simulations that require long readout windows.
While a beam spill at DUNE spans $\mathcal{O}(3~\mathrm{ms})$, set by the maximum electron drift time,
simulations of supernova neutrino bursts must cover time scales of order $100~\mathrm{s}$ to capture all relevant activity.
A supernova within our galaxy ($\sim \mathcal{O}(10)\,\mathrm{kpc}$) is expected to produce $\mathcal{O}(10^3)$
detectable neutrino interactions in the DUNE Far Detector~\cite{DUNE:2020zfm},
making full-window simulation with dense representations computationally challenging.
In this regime, the sparse, block-based approach of \texttt{TRED} offers a particularly attractive path toward scalable far-detector signal simulation.

\subsection{Truth information propagation in simulation}

We preserve the origin of the ionization charge by storing the mapping from 
each \texttt{G4Step} object identifier in the original edep-sim input
to the corresponding rasterized charge on each detector channel and short 
time slice.
This bookkeeping does not require additional GPU memory.
Its overhead comes from writing the sparse provenance record to output
and from assigning the corresponding \texttt{G4Step}
or trajectory identifiers to
rasterized charge contributions during runtime;
in the samples considered here,
the latter increases the processing time by approximately 5\%–10\%.
This sparse charge-based record supports truth labeling
without keeping a truth-decomposed history for every waveform sample,
which would be prohibitively expensive for highly segmented detectors.
For detectors with sampled readout, the stored truth charge can be directly
matched to the deconvolved signal in the corresponding channel-time region.

For the zero-suppressed ND-LAr readout, 
where the complete sampled waveform is not retained and standard
deconvolution is not directly applicable,
the stored provenance record is used through
an on-demand waveform-level truth-attribution procedure.
For a given reconstructed hit,
the rasterized charge contributions associated with the relevant
channel-time region are queried and propagated through the same
field-response convolution model used in the detector simulation.
The contribution from each \texttt{GEANT4} trajectory
is then evaluated within the start
and end time window of the reconstructed hit.
This procedure decomposes selected signals into trajectory-level
contributions and quantifies charge sharing between overlapping particles.
The waveform-level attribution adds approximately 80\% to the processing
time relative to the base simulation but requires no additional memory,
since the local detector response is recomputed from the saved provenance
rather than stored during the initial simulation.

The longer-term goal is to use the stored charge provenance
together with ongoing ND-LAr signal-processing and reconstruction
developments.
A complete truth-to-reconstruction matching study for the ND-LAr workflow,
which requires correcting for detector effects such as
thresholding and long-range induction
before the reconstructed quantities can be compared
to the original ionization charge,
is deferred to a dedicated reconstruction publication.

\subsection{Light signal simulation in LArTPC}
The present work focuses on the simulation of charge signals in liquid-argon TPCs.
Light simulation requires distinct treatment of optical-photon production,
propagation, and detection, which is computationally
and conceptually separate from ionization charge.
It is outside the scope of the current framework 
and is handled separately by dedicated tools~\cite{Blyth:2025qpe,Galgoczi:2025nic}.
This separation allows for a more modular design and easier maintenance,
and is also consistent with the treatment of experimental data,
where charge and light information are typically stored
and reconstructed as distinct data products.
A similar separation is common in modern LArTPC simulation frameworks,
including \texttt{larnd-sim}~\cite{larndsim}.
The correspondence between charge and light signals can nevertheless
be preserved through their common GEANT4 truth information,
which provides a shared reference for interfacing 
charge outputs with external light-simulation tools.

\section{Summary}\label{sec:summary}
In summary, we have presented \texttt{TRED}, a GPU-accelerated simulation package that enables accurate and efficient modeling of charge signal formation in liquid argon time projection chambers. The core of the approach is an analytic effective-charge formulation based on Gaussian quadrature, which captures sub-grid structure without requiring dense sampling. In conjunction with this, we introduce a generic block-sparse binned tensor representation that supports efficient reduction operations on sparse signals and enables the direct application of FFT-based algorithms to localized regions of the detector volume. Together, these components yield an FFT-based analytic solution for induced signals that demonstrates strong performance in both runtime and GPU memory usage under beam conditions and pile-up levels expected at DUNE–ND, while remaining applicable to other large-scale detectors with sparse activity, including DUNE–FD.

More broadly, the design principles and data abstractions introduced in \texttt{TRED} extend beyond a single detector or experiment. By combining analytic signal modeling with sparse, block-structured tensor representations and community-supported GPU software, this work presents a scalable and adaptable approach to detector simulation for experiments characterized by large volumes and highly localized activity. These concepts are readily transferable to other neutrino detectors and to computational domains that require efficient processing of large, sparsely activated datasets.

\section*{Acknowledgement}
We thank Matt Kramer, Dan Dwyer, Alex Booth, and Haiwang Yu for their helpful input on existing simulation frameworks.
We especially thank Matt Kramer for his thorough review of the manuscript.
We also acknowledge the use of ChatGPT for language editing.

\bibliographystyle{IEEEtran}
\bibliography{references}

\end{document}

%% file: simulation.tex
\section{Simulation Framework and Signal Formation}~\label{sec:overview_simulation}
\subsection{Detector physics and simulation workflow}~\label{sec:det}

In a LArTPC, charged particles traversing the detector deposit energy in the argon medium,
producing ionization electrons and scintillation light.
A fraction of the ionization electrons recombines with argon ions,
while the remaining free electrons drift toward the anode under an externally
applied electric field and induce signals on the readout electrodes.
During transport, drifting electrons may be attenuated by attachment to electronegative
impurities and they may undergo spatial diffusion.
These effects are accounted for in the simulation through parameterized models of recombination,
finite electron lifetime, and diffusion.

According to Shockley--Ramo’s theorem~\cite{Shockley1938,Ramo:1939},
the instantaneous current induced on an electrode by a moving charge $q$ is
\begin{equation}
  \label{eq:ramo}
I = -q\,\mathbf{v}_{\mathrm{drift}}\cdot\nabla W,
\end{equation}
where $\mathbf{v}_{\mathrm{drift}}$
is the charge drift velocity, determined by the external electric field
and the transport properties of liquid argon, and
$W$ is the weighting potential, which is fully defined by the detector geometry.
In general, the charge distribution evolves according to coupled
partial differential equations describing electrodynamic transport, attachment, and diffusion processes.

For typical LArTPC operating conditions, the transverse and longitudinal diffusion accumulated
over meter-scale drift distances remains limited to a few millimeters.
Since the induced signal can vary significantly only within a region of a few millimeters near the anode,
the evolution of the charge distribution during drift 
is assumed to be factorized from 
the signal induction process
following the procedure in Refs.~\cite{DUNE:2020ypp,DUNE:2022gxa}.
Diffusion effects are therefore incorporated through a static charge distribution evaluated near the anode plane,
while the induced current is computed assuming a unit point charge
following the nominal drift trajectory in the vicinity of the electrodes.
This factorization substantially simplifies the signal calculation while maintaining the accuracy
required for the detector configurations considered in this work.

Under this approximation, the total induced signal $\mathcal{I}(\mathbf{x}_p,t)$
on an electrode located at $\mathbf{x}_p$ is written as
\begin{equation}
\label{eq:greenfunction}
\mathcal{I}(\mathbf{x}_p,t) =
\int d^2\mathbf{x}' \int dt'\,
i(\mathbf{x}',t';\mathbf{x}_p,t)\, q(\mathbf{x}',t'),
\end{equation}
where $q(\mathbf{x}',t')$ denotes the charge distribution evaluated near the anode,
with diffusion effects already incorporated,
and $i(\mathbf{x}',t';\mathbf{x}_p,t)$ is the \textit{field response} function.
The field response function, which plays the role of a Green’s function for the signal induction problem,
is defined by evaluating Eq.~\eqref{eq:ramo} for a unit charge drifting along the nominal trajectory,
while neglecting attenuation and diffusion effects.

\begin{figure}[!t]
  \centering
  \scalebox{0.85}{%
  \begin{tikzpicture}[node distance=1.1cm]

  % Main pipeline (from Fig.1, with i/j annotations)
\node (start) [data] {G4Step objects (track-segments from GEANT4)};
    \node (seg) [process, below of=start] {Split long G4Step above threshold};
    \node (stepseg) [data, below of=seg] {Step segments};
    \node (p1) [process, below of=stepseg] {Apply charge transport [R;A;D]};
    \node (d1) [data, below of=p1] {Augmented step segments};

  % Raster stage (outer loop i)
  \node (p2) [process, below of=d1] {Rasterize to N-dimensional charge blocks [Q]};
  \node (d2) [data, below of=p2] {N-d charge blocks};
  \node (p3) [process, below of=d2] {Bin \& accumulate at same locations};
  \node (d3) [data, below of=p3] {Unique N-d charge blocks $Q(i)$};

  % Convolution stage (inner loop j)
  \node (p4) [process, below of=d3] {Convolve with field response [I]};
  \node (d4) [data, below of=p4] {Induced currents blocks $I(i,j)$};

  % Electrode aggregation (per i, then global)
  \node (p5) [process, below of=d4] {Bin \& accumulate on unique electrodes};
  \node (d5) [data, below of=p5] {Unique electrode currents $I(i)$};
  \node (p6) [process, below of=d5] {Electronic readout model};
  \node (d6) [data, below of=p6] {Data ready for analysis};

  % Arrows main line
  \draw [arrow] (start) -- (seg);
  \draw [arrow] (seg) -- (stepseg);
  \draw [arrow] (stepseg) -- (p1);
  \draw [arrow] (p1) -- (d1);
  \draw [arrow] (d1) -- (p2);
  \draw [arrow] (p2) -- (d2);
  \draw [arrow] (d2) -- (p3);
  \draw [arrow] (p3) -- (d3);
  \draw [arrow] (d3) -- (p4);
  \draw [arrow] (p4) -- (d4);
  \draw [arrow] (d4) -- (p5);
  \draw [arrow] (p5) -- (d5);
  \draw [arrow] (d5) -- (p6);
  \draw [arrow] (p6) -- (d6);

  % ----- Loop annotations (from Fig.2 control flow) -----

  % Inner loop: j over convolution for fixed i
  \coordinate (innerDown) at ($(d4)!0.5!(p5)$);
  \coordinate (innerUp)   at ($(d3)!0.5!(p4)$);
  \draw[arrow]
    (innerDown) -- ++(-42mm,0)
    node[pos=0.8, below]{next $j$}
    |- (innerUp);

  % Outer loop: i over raster+conv blocks
  \coordinate (outerDown) at ($(d5)!0.5!(p6)$);
  \coordinate (outerUp)   at ($(d1)!0.5!(p2)$);
  \draw[arrow]
    (outerDown) -- ++(-47mm,0)
    node[pos=0.8, below]{next $i$}
    |- (outerUp);

  \end{tikzpicture}
  }%
  \caption{
    Flowchart of signal formation and processing from G4Step objects (track segments produced by \textsc{GEANT4})
to analysis-ready simulation data. Physics processes are noted in brackets:
R: recombination at the initial ionization;
A: electron attachment during transport due to finite lifetime;
D: drift under the external electric field;
Q: rasterization and charge-weighting using Gaussian quadrature rules;
I: induced current calculation using the field response function.
The side arrows represent the control flow, annotating the outer loop over raster index $i$
and the inner loop over response index $j$. \label{fig:flowchart}}
\end{figure}

Figure~\ref{fig:flowchart} shows the full signal formation and processing pipeline.
Each \texttt{G4Step} (a track-segment in the \texttt{GEANT4} output) is described by its start and end points,
the total energy deposition, and the average energy loss per unit length.
These quantities are stored as a single row in a tensor-based representation used by the simulation framework.
In data-parallel processing, long and short \texttt{G4Step} objects are processed together.
When a long step is converted into voxels, it can create many empty voxels and waste computing resources.
To reduce this problem, we divide steps longer than a chosen threshold into smaller step-segments.
In this work, we use 5~mm as the threshold.
This value is close to the channel pitch in many LArTPCs and helps improve resource use.
Steps shorter than the threshold are kept unchanged.
The smaller step-segments keep the same $dE/dx$ as the original step, while the deposited energy is reduced in proportion to the segment length.
This preprocessing keeps the original \texttt{edep-sim} information and does not introduce bias.
Based on this initial step-segment representation, detector transport effects are applied to determine
the number of ionization electrons that survive recombination and attachment, as well as the characteristic
diffusion width accumulated during drift. These quantities are appended to the corresponding tensor row,
yielding an augmented step-segment representation.
Step segments are then mapped into the N-dimensional domain of the field response sampling, as described in Sec.~\ref{subsec:effq}.
For pixelized detectors, $\text{N}=3$: two transverse spatial dimensions and one temporal dimension along the drift paths.
This representation is chunked into N-dimensional tensors and the combination of one chunk
and the N-dimensional location of its lower corner is called a ``block''.
Blocks corresponding to identical spatial and temporal coordinates are summed to produce unique N-dimensional blocks.
The detector response is subsequently incorporated by convolving these blocks with the field response function,
yielding induced currents on the readout electrodes. All contributions are summed to produce a total induced current waveform for each electrode.
An electronic readout model is applied to produce signals suitable for higher-level physics analysis.

\subsection{Induced current modeling} \label{sec:induce-signal}
The LArTPC anode is composed of multiple electrodes,
whose geometry and arrangement differ between detector designs.
Owing to the translational symmetry of the electrode structure,
it is sufficient to evaluate the response at the $p$-th pixel,
and to derive the field response at the $(p+m)$-th pixel
by $i(\mathbf{x}';\mathbf{x}_{p+m}) \equiv i(\mathbf{x}';\mathbf{x}_{p}+m\mathbf{a}) = i(\mathbf{x}'-m\mathbf{a};\mathbf{x}_{p})$,\
where $\mathbf{x}_p$ denotes the position associated with pixel index $p$,
$p=(p_1,p_2)$ and $m=(m_1,m_2)$ are integer index pairs for the two-dimensional pixel array,
and $m\mathbf{a}$ denotes the lattice displacement $m_1\mathbf{a}_1 + m_2\mathbf{a}_2$
from the $(p+m)$-th pixel to the $p$-th pixel.
We use $i(\mathbf{r}')$ for the standard field response, when the central electrode is at the origin of the coordinate,
and therefore the field response at
$\mathbf{x}_p$ can be expressed by $i(\mathbf{x}';\mathbf{x}_p)=i(\mathbf{x}'-\mathbf{x}_p)$,
where $\mathbf{x}' = \mathbf{r}'+ \mathbf{x}_p$.
Along the drift path, the longitudinal coordinate $z$ is in one-to-one correspondence with time $t$.
The trajectory is integrated using a Runge–Kutta scheme, so that the time points corresponding to nominally constant time steps are obtained approximately.
Therefore, the three-dimensional source distribution can be represented as $q({\mathbf{r}}'+\mathbf{x}_p, t; t_{0})$,
and the field response written accordingly as $i({\mathbf{r}}', t-t_0)$, where ${\mathbf{r}}'=(x', y')$ denotes the coordinates within the pixel plane.
Equation~\eqref{eq:greenfunction} can be written as
\begin{equation}
    \mathcal{I}_p(t)\equiv\mathcal{I}(\mathbf{x}_p, t) =
    \int d^2 \mathbf{r}'\int dt' i(\mathbf{r}', t-t')q(\mathbf{r}'+\mathbf{x}_p,t'),
    \label{eq:firstapprox}
\end{equation}
where the electron is located at ${\mathbf{r}}'$ relative to the fixed electrodes.

\begin{figure}[t]
  \centering
  \includegraphics[width=0.95\linewidth]{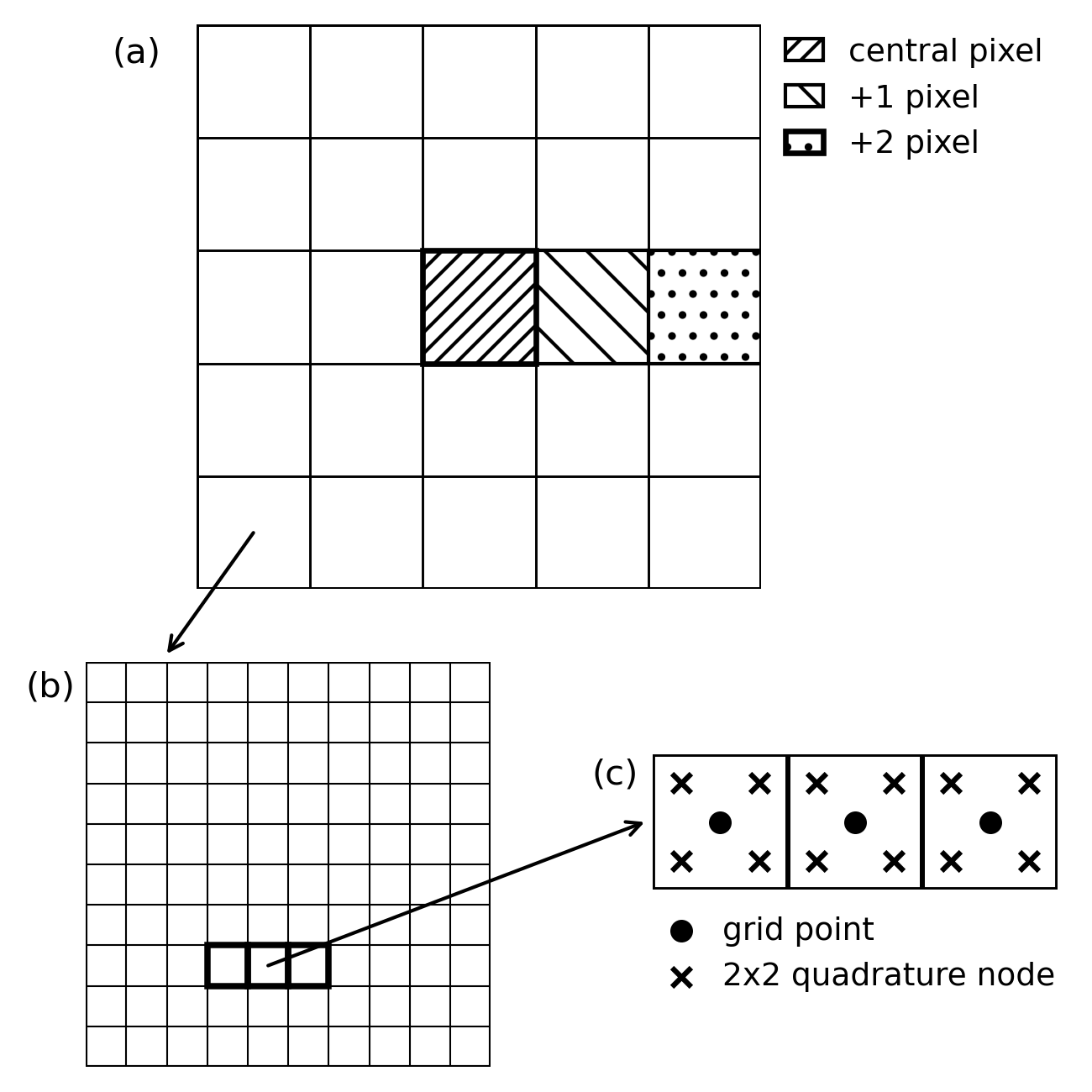}
  \caption{Illustration of the field response grid and quadrature
    nodes. (a) Field-response grid defined in pixel pitches, showing
    the central pixel of interest together with its neighboring pixel
    areas used in the response calculation.  (b) Enlargement of a
    single pixel area from (a), subdivided into a $10\times10$ grid of
    sub-cells, where electrons are released at the center of each
    sub-cell.  (c) Sampling within selected sub-cells, showing the
    electron release point (dot) and the associated quadrature nodes
    (crosses) used for the discrete formulation of the detector
    response in this article.}
  \label{fig:frillu}
\end{figure}

In practice, the field-response function $i(\mathbf{r}')$ is computed
by fixing the electrode geometry and evaluating the induced current on
a central electrode of interest for charges at different relative positions.
Since the induced signal depends only on the relative
geometry between the charge and electrodes, this is equivalent to
translating the electrodes past a fixed charge,
but is simpler and more efficient.
Figure~\ref{fig:frillu}(b) illustrates the sampling of a single pixel using 
$10 \times 10$ impact partition. Each impact position is located at the center
of a partitioned area within the pixel and defines
the starting location of a drift path over which the detector response is sampled.
Figure~\ref{fig:frillu}(c) shows the subsampling scheme within selected impact positions,
where the grid point
(dot) indicates the electron impact location used for sampling,
and the associated quadrature nodes (crosses) are shown for completeness;
their definition and role are discussed in Sec.~\ref{subsec:effq}.

% Figure~\ref{fig:frexample} shows an
% example of currents induced by a single electron drifting along selected paths above selected pixels of the DUNE ND-LAr pixelated anode:
% the top panel compares the center of the collection pixel and and a path near the border of the pixel,
% while the bottom panel shows the \(+1\) and \(+2\) neighboring pixels;
% the model was produced and validated with COMSOL and \textit{pochoir} packages~\cite{Martynenko_2023}.

% \begin{figure}[t]
%   \centering
%   \includegraphics[width=\linewidth]{figures/FRForTRED.pdf}
%   \caption{
% Induced current from a single drifting electron on the pixelated anode plane of the DUNE ND-LAr detector.
% Top: response measured on the central (reference) pixel for an electron located at the pixel center and at the pixel boundary.
% Bottom: responses measured on the first (+1) and second (+2) neighboring pixels for an electron located at the center of the reference pixel. The field response model is generated using COMSOL and validated with the \textit{pochoir} package.
%   }
%   \label{fig:frexample}
% \end{figure}

\begin{figure}
    \centering
    \includegraphics[width=\linewidth]{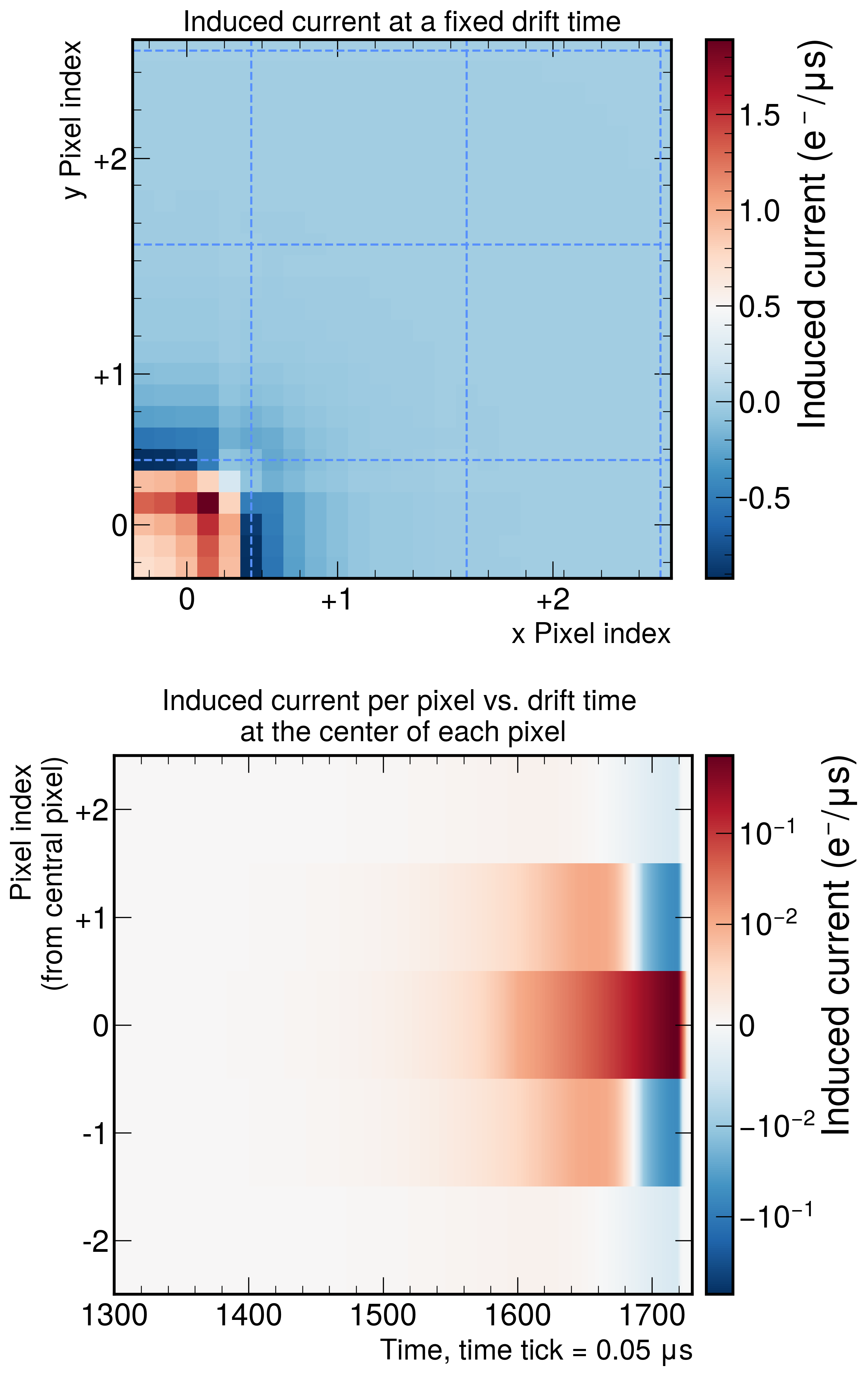}
    \caption{Spatial and temporal profiles of the induced current on the central and neighboring pixels. \textbf{Top}: Instantaneous 2D cross-sectional current induction at a fixed drift time near peak induction. \textbf{Bottom}: Time evolution of the induced current for the central pixel and its neighbors.}
    \label{fig:frfortred}
\end{figure}

Figure ~\ref{fig:frfortred} shows the current induced on the central and neighboring pixels in the DUNE ND-LAr pixelated anode. The top panel displays the instantaneous 2D cross-sectional current induction across the pixel plane at a fixed drift time near the peak induction, illustrating the spatial extent of the signal. The bottom panel shows the time evolution of the induced current along the drift path for the central and neighboring pixels. A symmetric logarithmic scale is applied to the bottom panel to highlight the low-amplitude signals on the outer pads. The field response is generated using the \textit{pochoir} package ~\cite{Martynenko_2023}.

\input{effq}
\input{convo}

\subsection{Readout electronics response}
\label{sec:readout}
There are two liquid argon in-situ charge readout electrode implementations:
projective readout and pixelated readout.
Each implementation has ramifications for electronics
readout system architecture design resulting in tradeoffs for signal extraction fidelity.
Both are supported by our simulation framework, and this work focuses on pixelated readout.

Projective readout is implemented as multiple offset planes of
arrays of parallel wire or strip electrodes at different angles,
providing redundancy for charge reconstruction.
Plane offset and wire/strip pitch are at the $\mathcal{O}$(1)\,mm-scale.
Anode co-planar coordinates are determined via
tomographic projections on all sense planes~\cite{Qian:2018qbv}.
Projective readout is typically operated
in a manner wherein current waveforms are continuously digitized at a fixed $\mathcal{O}$(1)\,MHz
sampling rate until a trigger initiates the readout of
a portion of waveforms spanning a pre-configured duration.
The heat from electronics is well controlled given the moderate number of channels inherent to the projective architecture.
The waveform on a channel is a superposition of the induced signal from drifting charges,
shaped by the cryogenic preamplifier,
front-end filters, and various forms of noise.
Several $\mathcal{O}$(10$^4$)-channel LArTPC detectors have operated with the projective
readout approach~\cite{MicroBooNE:2016pwy, SBND:2020scp, ICARUS-T600:2020ajz, DUNE:2021hwx}.

Pixelated readout is implemented as two dimensional arrays of conductive pixel
pads with $\mathcal{O}$(1)\,mm pixel pitch on large-format printed circuit boards.
Each pixel collects ionization electrons into a uniquely addressable
readout channel, enabling native three-dimensional imaging.
Each readout channel is an independent self-triggering signal processor capable of amplification,
digitization, and multiplexed readout, where $\mathcal{O}$(10) channels are digitally controlled
and configured with an application-specific integrated circuit (ASIC) digital core.
Freed ionization electron sparsity permits pixelated readout to nominally operate in a quiescent mode,
limiting power dissipation and enabling scaling to $\mathcal{O}$(10$^7$) channel systems~\cite{DUNE:2021tad}.

\begin{figure}[h]
  \centerline{\includegraphics[width=0.98\linewidth]{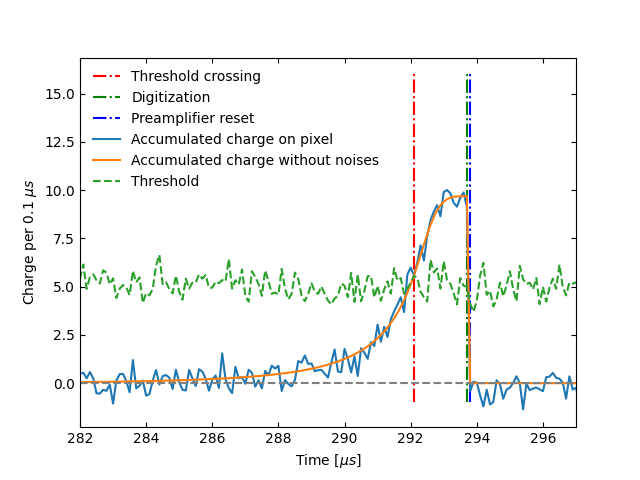}}
  \caption{Trigger--reset scheme of the LArPix ASIC~\cite{Dwyer:2018phu}.
    The channel threshold is represented by a horizontal dashed line.
  The vertical dashed line indicates the time at which the input signal crosses the discriminator threshold.
  The integrated charge on the preamplifier is sampled after a fixed window, marked by the dash--dotted line.
  Following a short reset interval, ending at the dotted line, charge accumulation on the preamplifier resumes.
  }\label{fig:readout}
\end{figure}

In the trigger-reset scheme~\cite{Dwyer:2018phu}, a digital record encodes
the triggered channel's timestamp and an ADC data word,
where the latter corresponds to the channel integrated charge at the time of threshold crossing
plus a configurable fixed integration duration and
the former marks time of digitization with a $\mathcal{O}$(10)\,MHz clock.
After the triggered readout, the channel is reset and briefly inactive.
An example of the trigger--reset readout sequence is shown in Fig.~\ref{fig:readout}.
This design is assumed throughout the rest of the work.
A variant trigger logic mode allows multiple, successive digitizations
to occur that avoid intervening channel resets~\cite{annrev:2024},
mimicking a continuously sampled readout for a fixed duration
to evade charge loss at channel reset.
In the fixed-charge integration scheme~\cite{Nygren:2018rbl},
timestamps of integrated fixed charge quanta are recorded and
the front-end charge is replenished without charge loss after each trigger.
Both schemes operate at the $\mathcal{O}(1)$\,kiloelectron channel threshold scale.
Charge bias and inefficiencies are incurred from charge
left on a channel analog front end due to threshold-based triggering.

%% file: effq.tex
\subsection{Effective charge in the simulation}
\label{subsec:effq}
In this section,
we discretize the continuous induced current in Eq.~\eqref{eq:firstapprox}
using an effective grid-based formulation that preserves fine spatial locality
without resorting to dense sampling.
The resulting integrals are evaluated using \textit{Gauss--Legendre quadrature rule}, i.e.,
\[
  \int_{-1}^{1} \rd x \, h(x) \approx \sum_i w_i h(\xi_i),
\]
where $\{\xi_i, w_i\}$ are the nodes and weights of the $N$-point
Gauss--Legendre rule~\cite{Gautschi1970}.
In practice, the field response function is evaluated on an equally spaced grid $G$.
We therefore express the induced current as a discrete summation over grid points
by introducing an effective charge $Q^{\mathrm{eff}}$ on the grid,
\begin{equation}
\mathcal{I}(\mathbf{x}_p,t) =
\sum_{g \in G} Q^{\mathrm{eff}}(\mathbf{r}_g,t_g)\,
I(\mathbf{r}_g,t_g;\mathbf{x}_p,t),
\label{eq:iqsum}
\end{equation}
where $I(\mathbf{r}_g;\mathbf{x}_p)=i(\mathbf{r}_g-\mathbf{x}_p)$ is the field response
evaluated at the grid point $\mathbf{r}_g$
in the pixel plane and at the corresponding drift time $t_g$.
This formulation enables Gaussian quadrature on a regular grid,
even when the quadrature nodes are not aligned with the grid points.
For notational simplicity, the explicit time dependence is suppressed in the following.

We now derive the effective charge by partitioning the detector volume into
cuboidal elements whose dimensions match the granularity of the field response function,
establishing a one-to-one correspondence with the grid $G$.
The continuous integral can then be written as
\begin{align}
  \int \rd \mathbf{r}' \, Q(\mathbf{r}') I(\mathbf{r}';\mathbf{x}_p) 
  &= \sum_{C} \int_{C}\rd \mathbf{r}'_{C} \, Q(\mathbf{r}'_{C}) \, I(\mathbf{r}'_{C};\mathbf{x}_p). \label{eq:sumcub}
\end{align}
where the sum runs over all cuboids $C$.
Applying Gaussian quadrature locally within each cuboid yields
\begin{align}
 \int_{C} Q(\mathbf{r}') I(\mathbf{r}') \, \rd \mathbf{r}' 
 &\approx \sum_{l,m,n}  w_{lmn} \, Q_{C,lmn} \, I_{C,lmn}, \label{eq:cub}
\end{align}
where indices $(l,m,n)$ denote the quadrature nodes inside cuboid $C$.

To evaluate the field response at a quadrature node,
we interpolate from the field response function evaluated at nearby grid points.
For a quadrature node within a cuboid $C$, the field response is obtained
by trilinear interpolation using the response at the center of $C$ and its
twenty-six neighboring cuboids,
\begin{align}
  I_{C,lmn} &= \sum_{r,s,t} I_{C,rst} \, u_{lmn,rst}, \label{eq:trilinear}
\end{align}
where $(r,s,t) \in \{-1,0,1\}^3$ denote the cuboid itself ($0$)
and its lower ($-1$) and upper ($1$) neighboring cuboids along each axis,
and $u_{lmn,rst}$ are the piecewise trilinear interpolation coefficients
determined by the relative position of the quadrature node within the cuboid.

Substituting Eqs.~\eqref{eq:cub} and~\eqref{eq:trilinear} into
Eq.~\eqref{eq:sumcub} yields
\begin{equation*}
  \label{eq:rearr}
  \begin{aligned}
  \mathcal{I}(\mathbf{x}_p) 
  &= \sum_{C} \sum_{l,m,n}  w_{lmn} \, Q_{C,lmn}  
     \left( \sum_{r,s,t} I_{C,rst} \, u_{lmn,rst} \right) \\
  &= \sum_{C}\sum_{r,s,t} I_{C,rst} 
     \left( \sum_{l,m,n} Q_{C,lmn} \, w_{lmn} \, u_{lmn,rst} \right).
\end{aligned}
\end{equation*}

Each cuboid $C$ is labeled by indices $(i,j,k)$ along each axis.
Together with its neighbors, it spans the set $(i+r,j+s,k+t)$  with $(r,s,t) \in \{-1,0,1\}^3$. 
Combining the summations over cuboids and their neighbors allows Eq.~\eqref{eq:rearr}
to be recast in terms of grid points $(i,j,k)$,
\begin{equation}
\label{eq:effq}
\mathcal{I}(\mathbf{x}_p) 
= \sum_{i,j,k} Q_{ijk}^{\text{eff}} \, I_{ijk},
\end{equation}
where the effective charge is defined as
\begin{equation}\label{eq:eff_q}
Q_{ijk}^{\text{eff}} = \sum_{l,m,n} \sum_{r,s,t} w_{lmn} \, Q_{lmn,i-r,j-s,k-t} \, u_{lmn,rst}.
\end{equation}
The effective charge $Q_{ijk}^{\text{eff}}$ thus corresponds to a discrete convolution
of the ionization charge at quadrature nodes $(l,m,n,i,j,k)$ with the trilinear
interpolation coefficients $u_{lmn,rst}$ along the last three indices.

\subsubsection{Calculation of $Q$ for a line-shape step segment}
\label{sec:chargedist}
An ionized track is represented by a series of line segments defined by their two endpoints:
\[
  \mathbf{p_0} = (x_0, y_0, z_0) \quad \text{and} \quad \mathbf{p_1} = (x_1, y_1, z_1).
\]
Diffusion is characterized by the coefficients  \(\sigma_x, \sigma_y, \sigma_z\) in the $x$, $y$, and $z$
directions, respectively.
The resulting
ionization charge distribution is given by the convolution of a
uniform line charge with a three-dimensional Gaussian diffusion
kernel, and it admits a closed-form expression.:
  \begin{equation}
    \begin{aligned}
    q(x, y, z) &=  \, Q \frac{\exp(-B/\Delta^2)}{4\pi\Delta} \times \\
    &\biggl(\erf \bigl(\frac{A_2}{\sqrt{2} \Delta \sigma_x \sigma_y \sigma_z} \bigr) - 
      \erf\bigl(\frac{A_1}{\sqrt{2} \Delta \sigma_x \sigma_y \sigma_z} \bigr) \biggr)
      \end{aligned}
      \label{eq:qline}
  \end{equation}
where
\begin{align*} \label{eq:denumdelta}
    \Delta^2 = &\, \sigma_y^2 \sigma_z^2 (x_0 - x_1)^2 + \sigma_z^2 \sigma_x^2 (y_0 - y_1)^2 + \sigma_x^2 \sigma_y^2 (z_0 - z_1)^2,\\
    \begin{split}
    A_1 = &\, \sigma_y^2 \sigma_z^2 (x-x_0)(x_0 - x_1) + \sigma_z^2 \sigma_x^2  (y-y_0)(y_0-y_1) \\
       & + \sigma_x^2  \sigma_y^2 (z-z_0) (z_0 -z_1),
    \end{split} \\
    \begin{split}
    A_2 = &\, \sigma_y^2 \sigma_z^2 (x-x_1)(x_0 - x_1) + \sigma_z^2 \sigma_x^2  (y-y_1)(y_0-y_1) \\
       & + \sigma_x^2  \sigma_y^2 (z-z_1) (z_0 -z_1), 
    \end{split} \\
    \begin{split}
    B = &\, \sigma_y^2 \bigl( x (z_0 - z_1) + (z_1 x_0 - z_0 x_1) - z (x_0 - x_1) \bigr)^2 \\
       &+ \sigma_z^2 \bigl( y (x_0 - x_1) + (x_1 y_0 - x_0 y_1) - x (y_0 - y_1) \bigr)^2 \\
       &+ \sigma_x^2 \bigl( y (z_0 - z_1) + (z_1 y_0 - z_0 y_1) - z (y_0 - y_1) \bigr)^2,
    \end{split} \\
    \erf \coloneq &\,\text{error function}.
\end{align*}
Accordingly, the discrete charge contribution at the quadrature node $(l,m,n)$ within grid cell $(i,j,k)$ can be written as
\[
Q_{lmn,ijk} = q(x_{li},\, y_{mj},\, z_{nk}).
\]
\subsubsection{Calculations of $Q$ for short step segments} 
In the short-segment limit, Eq.~\eqref{eq:qline} reduces to a point-like Gaussian
distribution centered at the segment midpoint.
Since the factor $\Delta$ appears in the denominator of
Eq.~\eqref{eq:qline}, this limit can lead to numerical instabilities.
To ensure numerical stability, we switch between the point-like and
line-shaped formulations: if all components of
$\mathbf{p}_1-\mathbf{p}_2$ are smaller than $0.1\,\sigma$ (configurable)
along the corresponding axes, the point-like formulation is used;
otherwise, the line-shaped formulation is applied.

%exact code ... did not check if things are consistent, but the coding should follow the bottom one ...
%"(-Erf((Power(sx,2)*Power(sz,2)*(y - y0)*(y0 - y1) + Power(sy,2)*(Power(sz,2)*(x - x0)*(x0 - x1) + Power(sx,2)*(z - z0)*(z0 - z1)))/(Sqrt(2)*sx*sy*sz*Sqrt(Power(sx,2)*Power(sz,2)*Power(y0 - y1,2) + Power(sy,2)*(Power(sz,2)*Power(x0 - x1,2) + Power(sx,2)*Power(z0 - z1,2)))))+ Erf((Power(sx,2)*Power(sz,2)*(y - y1)*(y0 - y1) + Power(sy,2)*(Power(sz,2)*(x - x1)*(x0 - x1) + Power(sx,2)*(z - z1)*(z0 - z1)))/(Sqrt(2)*sx*sy*sz*Sqrt(Power(sx,2)*Power(sz,2)*Power(y0 - y1,2) + Power(sy,2)*(Power(sz,2)*Power(x0 - x1,2) + Power(sx,2)*Power(z0 - z1,2))))))/(4.*Power(E,(Power(sz,2)*Power(x1*(-y + y0) + x0*(y - y1) + x*(-y0 + y1),2) + Power(sy,2)*Power(x1*(-z + z0) + x0*(z - z1) + x*(-z0 + z1),2) + Power(sx,2)*Power(y1*(-z + z0) + y0*(z - z1) + y*(-z0 + z1),2))/(2.*(Power(sx,2)*Power(sz,2)*Power(y0 - y1,2) + Power(sy,2)*(Power(sz,2)*Power(x0 - x1,2) + Power(sx,2)*Power(z0 - z1,2)))))*Pi*Sqrt(Power(sx,2)*Power(sz,2)*Power(y0 - y1,2) + Power(sy,2)*(Power(sz,2)*Power(x0 - x1,2) + Power(sx,2)*Power(z0 - z1,2))))"

%% file: convo.tex
\subsection{Interlaced convolution in calculating induced signals}
It is useful to rewrite Eq.s~\eqref{eq:firstapprox} and~\eqref{eq:iqsum} as
\begin{equation*}
\mathcal{I}_p(\mathbf{x}_p, t) = \sum_{g \in G} Q^{\text{eff}}(\mathbf{r}_g+\mathbf{x}_p, t_g)i(\mathbf{r}_g, t-t_g),
\end{equation*}
which is a cross-correlation in the pixel plane and a convolution along the time direction.

The pixel index $p$ corresponds to a coarser spatial grid with
spacing $\mathbf{a}$ than the finer grid $G$.
The finer grid arises from sampling multiple electron impact positions
within a single pixel when evaluating the field response function,
and thus these impact positions are naturally part of the grid $G$.
Consequently, the set of pixel indices $p$ forms a subset of the grid indices $g$ in the spatial domain;
equivalently, the pixel grid can be viewed as a supergrid of $G$.

We further decompose the grid index into a pair $(j,k)$,
where $k$ labels the pixel and $j$ indexes the sub-pixel grid points,
corresponding to the offset $\bm{\lambda}_j$ within that pixel.
Writing
\(
\mathbf{r}' = \bm{\lambda}_j + k \mathbf{a},
\)
we obtain
\begin{equation}
\begin{split}
\mathcal{I}_p
&= \sum_j \sum_k
Q^{\mathrm{eff}}(\bm{\lambda}_j + k \mathbf{a} + \mathbf{x}_p)\,
i(\bm{\lambda}_j + k \mathbf{a}) \\
&\equiv \sum_j \sum_k
Q^{\mathrm{eff}}[j, p+k]\,
i[j, k],
\end{split}
\end{equation}
after suppressing the explicit time dependence.

The cross-correlation in the pixel domain and the convolution in the time domain
can be further unified so that acceleration by low-level libraries and hardware
can be fully exploited.
This unified approach avoids the overhead associated with treating different dimensions separately.
We relabel the field response by
\(
R[j,k] \equiv i[j,N-k],
\) where $N$ is the length of the field response array.
This approach allows the expression
\begin{equation}
\mathcal{I}_p
=
\sum_j \sum_k
Q^{\mathrm{eff}}[j, p-k]\,
R[j, k].
\label{eq:interlaced}
\end{equation}
in which all dimensions appear in a convolution-like structure.
For the LArTPC geometry considered here,
the field response is symmetric about the collection electrode,
and our simulation exploits this property to further accelerate the package,
as discussed in Sec.~\ref{sec:mpcp}.
However, this symmetry is not required by Eq.~\eqref{eq:interlaced} itself.
We refer to the special treatment in Eq.~\eqref{eq:interlaced}
as an \emph{interlaced convolution}.
It should be noted that the effective charge in Eq.~\eqref{eq:effq}
is transformed from the grid $G$ to a representation in terms of the pixel grid
and its sub-pixel offsets in Eq.~\eqref{eq:interlaced}.

Finally, we note that within the discrete formalism
the grid points $g$ need not be uniformly spaced.
Any set of grid points may be employed, provided they can be mapped to integer indices.
This can be useful to more finely sample the field response
in regions near pixel borders where it varies more than pixel centers.
Variations in grid spacing are naturally absorbed into the quadrature weights $\mathbf{w}$.
In this sense, the discrete convolution introduced here constitutes
a generalized numerical formulation.